\documentclass{article}
\usepackage{psfrag}
\usepackage{graphicx}
\usepackage{color}
\usepackage{amsmath}
\usepackage{epsfig}
\usepackage{amssymb,amsfonts}
\usepackage{stmaryrd}

\begin{document}


\author{Pierre Bieliavsky\thanks{bieliavsky(AT)math.ucl.ac.be,
 D\'epartement de Math\'ematique  Universit\'e Catholique de Louvain,
 Chemin du Cyclotron 2,
1348 Louvain-la-Neuve, Belgium}, \, Charles Jego\thanks{
charles.jego(AT)cpht.polytechnique.fr, Centre de Physique Th\'eorique,
 Ecole Polytechnique, CNRS, 91128 Palaiseau Cedex, France, Unit{\'e}
 mixte de recherche du CNRS, UMR 7644. Preprint: CPHT-RR066.0906
}
\, and  Jan Troost\thanks{Jan.Troost(AT)lpt.ens.fr, Laboratoire de Physique
  Th\'eorique de l'\'Ecole Normale Sup\'erieure, 24 Rue Lhomond  75231 Paris
  Cedex  05, France,
 Unit{\'e} mixte
  du CNRS et de l'Ecole Normale Sup{\'e}rieure associ\'ee \`a l'universit\'e
Pierre et Marie Curie, UMR 8549. Preprint:  LPTENS-06/45}}

\title{Open strings in Lie groups and associative products}

\maketitle

\abstract{Firstly, we generalize a semi-classical limit of open
  strings on D-branes in group manifolds. The limit gives rise to
  rigid open strings, whose dynamics can efficiently be described in
  terms of a matrix algebra. Alternatively, the dynamics is coded in
  group theory coefficients whose properties are translated in a
  diagrammatical language. In the case of compact groups, it is a
  simplified version of rational boundary conformal
  field theories, while for non-compact groups, the construction gives
  rise to new associative products.  Secondly, we argue that the
  intuitive formalism that we provide for the semi-classical limit,
  extends to the case of quantum groups. The associative product we
  construct in this way is directly related to the boundary vertex
  operator algebra of open strings on symmetry preserving branes in
  WZW models, and generalizations thereof, e.g. to non-compact groups.
  We treat the groups $SU(2)$ and $SL(2,\mathbb{R})$ explicitly.  We
  also discuss the precise relation of the semi-classical open string
  dynamics to Berezin quantization and to star product theory.}

\newpage

\tableofcontents

\section{Introduction}
Theories of quantum gravity are expected to be non-local. String theory,
a prominent
theory of quantum gravity, is indeed non-local. A
non-commutative subsector of string theory has been identified that
decouples from gravity, but that retains some of the non-locality of
string theory. (See e.g. \cite{Seiberg:1999vs} and references therein,
and \cite{Douglas:2001ba} for a review.)  That discovery gave rise to
the further study of non-commutative field
 theories \cite{Snyder:1946qz}, and the
associated mathematical structures.  One can be hopeful that the study
of these theories  will teach us more about how to sensibly
formulate non-local theories of quantum physics and non-local
theories of quantum gravity.

In this article, we study a class of non-commutative theories.
Our study is motivated by the physics of open strings on D-branes in
group manifolds, but will not be restricted to this setting only, and
it will be explained in elementary and general terms. For open strings
on group manifolds, the above limit, decoupling from gravity, but
keeping highly non-trivial (non-commutative) gauge theory dynamics is
also available. It gives rise to the study of non-commutative gauge
theories on curved manifolds (see e.g. \cite{Schomerus:1999ug}).

In particular, we will firstly concentrate on a semi-classical limit
of open strings on generic group manifolds. This will have the
advantage of considerably simplifying the analysis of the open string
dynamics (while sacrifying finite bulk curvature effects).  This will
allow us to provide an elegant semi-classical picture of open string
dynamics on group manifolds. In order to present it, and to show its
generality, we delve into the theory of quantization of co-adjoint
orbits (which describes the behavior of one end of an open string),
the quantization of pairs of orbits (for the two endpoints of an open
string) and the composition of operators on the resulting Hilbert
spaces (corresponding to the concatenation of open strings).

For compact Lie groups, this part of our analysis is a simpler,
limiting version of the analysis of chiral classical conformal field
theory \cite{Moore:1988qv}, or boundary vertex operator algebras
\cite{Cardy:1991tv}.  We believe that providing an intuitive and
precise picture for the semi-classical limit of rational boundary
conformal field theory is interesting in its own right, and we
furthermore show that the picture also applies to non-rational
boundary conformal field theory which is much less understood.  Thus,
we are able to put our construction to good use, by explicitly
applying it to the example of non-compact groups. Moreover, it will
enable us to provide a proposal for a mathematical program quantizing
symplectic leaves of Poisson manifolds, and providing a product of
operators associated to triples of symplectic leaves.

Secondly, we argue that our intuitive picture that holds for
quantization for orbits of Lie groups, can be extended to quantum
groups, and that in doing so, we can recuperate the full solution to
rational boundary conformal field theories in the case of symmetry
preserving D-branes in Wess-Zumino-Witten models on compact groups,
and that the construction will generalize to non-compact
groups.
 
Since we make reference to a lot of concepts that have been highly
developed by different communities, we are not able to 
 review them all fully.
Our strategy will be to first illustrate the concepts in the simple
case of the group $SU(2)$. We then sketch how to obtain much more general
results on these
structures by citing the relevant references. 
In this way, we hope to show generality
while preserving intuition. We follow this strategy in section
\ref{orbitmethod} that reviews the orbit method, in section
\ref{pairs} that generalizes it to pairs of orbits, in section
\ref{su2} that reviews the structure constants for the $SU(2)$
case, as well as in appendix \ref{berezin} where we review some facts
concerning Berezin quantization and star product theory. Section
\ref{pairs} introduces and analyzes the action for rigid open strings
on group manifolds, and in sections \ref{sl2rbis} and \ref{noncompact}
we work out how
our formalism applies to the non-compact group $SL(2,\mathbb{R})$.
Finally, we treat the case of quantum groups in section
\ref{quantumgroups}.
\section{Co-adjoint orbit quantization}
\label{orbitmethod}
\begin{figure}
\centering
\includegraphics[width=5.5cm]{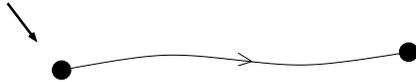}
\caption{We concentrate first on one endpoint of an open string.}
\label{oneendpoint}
\end{figure}

We start building an elementary picture of open string interactions by
concentrating on one of the endpoints of the open string (see figure
\ref{oneendpoint}).
Using the mathematical framework of the orbit method, we quantize the
associated degrees of freedom \footnote{We will take the liberty of
  always loosely referring to the intuitive picture of open strings.
  Most of the models we discuss can indeed be embedded in open superstring
  theory. However, a large part of our analysis applies in a
  more general context.}.
\subsection{An electric charge in the presence of a magnetic monopole}
In this subsection, we analyze the quantization of a charged particle
on a sphere, with a magnetic monopole sitting at the center of the
sphere. The particle will only interact electromagnetically. We show
that the quantization of the particle boils down to the quantization
of a co-adjoint orbit of $SU(2)$. 
\subsubsection*{The magnetic monopole}
We consider the three-dimensional flat space $\mathbb{R}^3$, with
spherical coordinates:
\begin{eqnarray}
x^1 &=& r \sin \theta \cos \phi
\nonumber \\
x^2 &=& r \sin \theta \sin \phi
\nonumber \\
x^3 &=& r \cos \theta
\end{eqnarray}
We put a magnetic monopole of integer charge $n$ at the origin of the
coordinate system $x^i=0$. We can describe the magnetic field it
generates in terms of the vector potentials:
\begin{eqnarray}
A^{\pm} &=& \frac{n}{2} (\pm 1 - \cos \theta) d \phi
\nonumber \\
F&=& d A^{\pm} = \frac{n}{2} \sin \theta d \theta \wedge d \phi
\nonumber \\
A^+ &=& A^- + n d \phi = A^- + d( n \phi)
\end{eqnarray}
where the vector potentials are valid near north ($\theta=0$) and
south ($\theta = \pi$) poles respectively (since we avoid a conical
singularity there due to the $\pm 1$ in the potential), and where the
relation between the two potentials is a gauge transformation on the
overlap, which is well-defined provided $n$ is indeed integer.

\subsubsection*{A charged particle on the sphere}
\begin{figure}
\centering
\includegraphics[width=5.5cm]{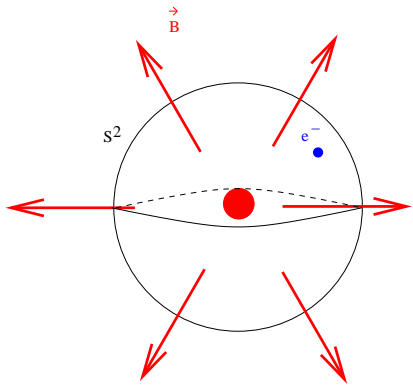}
\caption{An electron bound to a sphere, with a magnetic monopole in the middle
  of the sphere.}
\label{magnmonandcharge}
\end{figure}

Next, we introduce a charged particle, and we constrain it on a sphere
in the presence of the magnetic monopole (see figure
\ref{magnmonandcharge}).
As an action, we take the electromagnetic coupling only
\cite{Nielsen:1987sa,Alekseev:1988vx}:
\begin{eqnarray}
\label{san}
S =
\int A^{\pm} = 
\frac{n}{2}   \int d \tau (\pm 1 - \cos \theta) \dot{\phi}
\end{eqnarray}
The equations of motion are easily derived, and they imply that the
velocity of the particle is zero. The solutions to the equations of
motion thus correspond to the particle sitting at a fixed point of the
two-sphere. Thus, the classical phase space, which is the space of
classical solutions, is the two-sphere.

The global symmetries of the problem at hand are the $SU(2)$ rotations
which act transitively on the phase space.  The charges associated to
the $SU(2)$ global symmetries are the positions $x^i$ of the particle,
which are indeed conserved (since the particle does not move). They
can be shown to satisfy $SU(2)$ commutation relations under the Dirac
bracket (see e.g.  \cite{Balachandran:1983pc}). The Hamiltonian of the
purely topological action $(\ref{san})$ is zero.

\subsection{A classical particle on a co-adjoint orbit}
To generalize the well-known facts discussed above, we observe that
the gauge potential can be viewed as arising from the following
general formula for a one-form defined on a group manifold:
\begin{eqnarray}
\theta_\lambda (g) &=& \langle \lambda,g^{-1} dg \rangle
\end{eqnarray}
where $g$ is a group element of a Lie group $G$, $\lambda$ is an
element of the dual of the Lie algebra ${\cal G}^{\ast}$ and the
bracket $\langle .,. \rangle$ denotes the evaluation of $\lambda$ on
the element of the Lie algebra $g^{-1} dg$. Now that we
have defined the 
one-form, we can pull it back onto the world-line $L$ of a
particle via the map of the particle into the group manifold $g: L
\mapsto G : \tau \mapsto g(\tau)$, and we can define the action of
the particle as the integral over the world-line of this
one-form. It is an easy calculation to show that this action for the
case of $SU(2)$ precisely coincides with the action introduced above
\cite{Alekseev:1988vx}.  However, we can now generalize this action to
any group manifold, once we are given an element $\lambda$ of the dual
Lie algebra \cite{Alekseev:1988vx}:
\begin{eqnarray}
S &=& \int_{L} d \tau <\lambda,g^{-1} \partial_\tau g>
\end{eqnarray}
Again, the Hamiltonian corresponding to the action is zero. It is
interesting to analyze the symmetries of this action. The global
symmetry is $G$ which acts on the particle trajectory $g$ by
multiplication on the left. The local symmetry, i.e. the gauge group,
is the stabilizer of the weight $\lambda$, and it acts by
multiplication on the right. The local symmetry makes for the fact
that the particle is interpreted not as moving on the full group
manifold $G$, but rather on the manifold $G/Stab(\lambda)$ (which for
the case of $SU(2)$ is the two-sphere $SU(2)/U(1)=S^2$), which
coincides with the phase space.

The conserved Noether charges 
associated to the global symmetry group are $I=g
\lambda g^{-1}$ and they satisfy the Dirac brackets with the structure
constants equal to the structure constants of the Lie algebra of the
group \cite{Balachandran:1983pc}. (They are the generalization of the
positions $x^i$.) The symmetry group $G$ acts transitively on the phase
space. The global charges and functions thereof are gauge invariant
observables of the theory.

\subsection{The quantization}
In this subsection,
we quantize
the classical theory, starting out with
the $SU(2)$ example before generalizing to other groups.
\subsubsection*{The quantization of the spherical phase space}
We can quantize the phase space to find the Hilbert space. The
dimension of the Hilbert space is the number of Planck cells that fit
into the two-sphere. The symplectic form arising from the action
is the volume form of the two-sphere with quantized overall
coefficient, and consequently (after a straightforward calculation)
the number of Planck cells in phase space
is computed to be the integer number $n$. Since the group
$SU(2)$ is represented on the Hilbert space, we find a spin
$j=\frac{n-1}{2}$ representation of $SU(2)$ as the Hilbert space of
the particle. The group acts transitively on the classical phase
space, and is represented irreducibly on the quantum Hilbert space.

\begin{figure}
\centering 
\includegraphics[width=5cm]{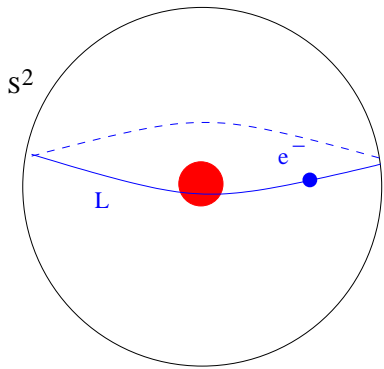}
\caption{The (blue) curve traced by the electron associated to a Wilson loop.
  The Wilson loop can be computed by calculating the flux through the
  upper cap (-- the hat --) or the lower cap (-- the bowl --), and
  must be unambiguous.}
\label{wilsonloop}
\end{figure}

We note that the (closed) Wilson loop $\exp (\imath \int A)$ (which in
the quantum theory weighs the path integral) and the quantization of
the coefficient of the action (which is the magnetic monopole charge)
can also be obtained by demanding that the path integral be
well-defined for closed world-line trajectories.  
Namely, the action should not be ambiguous, up to a
multiple of $2 \pi \imath$. Since the action can be computed by
calculating the flux either through the cap or the bowl (i.e. filling
in the Wilson loop (see figure~\ref{wilsonloop}) either over the north
or the south pole), we must have that the difference (divided by $ 2
\pi \imath$) is quantized.
Thus, the (half-integer) quantization of spin $j$ in this context is
re-interpreted in the electron/magnetic monopole system as being
associated to the Dirac quantization condition for the product of
electric and magnetic charge $n$. Also, invariance of the action up to
a multiple of $2 \pi \imath$ (which is the condition of
well-definedness of the quantum path integral), corresponds in the
geometric language to the fact that the two-sphere needs to be
integral \cite{Kirillov}.  Alternatively, in geometric quantization,
the $U(1)$ bundle over the two-sphere needs to be well-defined. All
these features are explained in terms of physics and fiber bundles in
\cite{Balachandran:1983pc} where the above constrained system is also
canonically quantized. The path integral approach to quantization was
developed in references \cite{Nielsen:1987sa,Alekseev:1988vx}.
Different regularizations of the path integral exist (see e.g.
\cite{Alekseev:1988vx} and the appendix of \cite{Troost:2003ge}) which
can also be understood as different types of geometric quantization
(namely naive, or metaplectic geometric quantization in the
nomenclature of \cite{Woodhouse}). These different regularizations
give slight shifts in the interpretation of the coefficient of the
action (as twice the spin or rather $2j+1$). (In the mathematics
literature as well (see e.g. \cite{Kirillov}) this ambiguity is
well-known.)


\subsection{The orbit method}
We very briefly review the highly developed domain of co-adjoint orbit
quantization. See e.g. \cite{Kirillov} for a summary.

\begin{figure}
\centering 
\includegraphics[width=5cm]{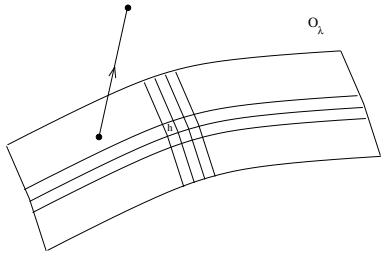}
\caption{The quantized orbit $O_\lambda$, on which the open string
endpoint lives, is divided into phase
space cells of size $h$ where $h$ is Planck's constant.}
\label{orbitquantization}
\end{figure}

The orbit method constructs the irreducible representations of a Lie
group $G$ via the study of its co-adjoint orbits. A co-adjoint orbit
$O_\lambda$ can be defined as the set obtained from a fixed
representative $\lambda$, an element of the dual of the Lie algebra of
the group, by the co-adjoint action of the group (i.e. by conjugation with
a group element). A co-adjoint orbit
has a canonical symplectic form $\omega_\lambda$, which can locally be
described as the derivative of the one-form $\theta_\lambda$ we
introduced above:
\begin{eqnarray}
\omega_\lambda &=& d \theta_\lambda
\end{eqnarray}
To quantize the orbit, we need the integral of the symplectic form
over closed two-cycles to be an integer -- this is the generalization
of the Dirac quantization condition. Orbits satisfying this condition
are called integral orbits. The orbits can then be quantized (see
figure~\ref{orbitquantization})
and each 
orbit gives rise to a Hilbert space which is 
an irreducible representation of the group
$G$.

A point we wish to stress here is that this construction is very
generic. When applied to $SU(2)$ it gives the results above, and when
applied to compact groups, one recuperates all irreducible
representations of the compact group, via the Borel-Weyl-Bott theorem
(see e.g. \cite{Kirillov}).
But we note that the orbit method has been successfully applied to a
much larger class of groups, including nilpotent groups, exponential
groups, solvable groups in general, as well as other finite
dimensional non-compact groups.  Moreover, it has been successfully
applied to infinite dimensional groups as well, like the Virasoro
group and loop groups.  We refer to \cite{Kirillov} for an extended
list of references and also to \cite{Taylor:1993zp} for many
references within the physics literature.\footnote{ It should be
  remarked that the orbit method does not necessarily give all
  irreducible unitary representations of a given group. In particular
  for finite dimensional groups, it typically gives rise only to those
  representations appearing in the decomposition of the left-regular
  representation of the group on the quadratically integrable
  functions. This will be important for us later, when we discuss the
  representations appearing in a particular non-compact example.}
\section{Pairs of orbits}
\label{pairs}
In this section, we discuss the addition of a second charged particle
to our physical problem. It corresponds to the second endpoint of the
open string (see figure~\ref{oneendpoint}). Firstly, we discuss
briefly the case where the two endpoints do not interact, and then we
discuss an interaction term which coincides with the Hamiltonian for
rigid open strings.
\subsection*{Tensoring for two}
We add a second particle and define in a first step its action to be
again the purely electromagnetic coupling, but with the opposite sign
for the charge (or equivalently, the opposite sign symplectic form).
We can add this particle to the same orbit as the first particle, but
we can also choose to constrain it to its own orbit. After
quantization, we find that the combined two-particle system has a
Hilbert space which is the tensor product of the $\lambda_I$
representation of the first particle and the $\bar{\lambda_F}$
representation of the second particle. (The indices $I,F$ indicate the
initial and the final endpoints of the string, respectively. The bar
on $\lambda_F$ reminds us that we quantize with the opposite sign
symplectic form, giving rise to the conjugate representation.)

The action for the uncoupled system is simply the sum of the two individual
particle actions:
\begin{eqnarray}
S_{2-particle} &=& \int_{L} d \tau <\lambda_I,g^{-1}_I \partial_\tau g_I>
     - \int_{L} d \tau <\lambda_F,g^{-1}_F \partial_\tau g_F>
\end{eqnarray}
The global symmetry is doubled to $G_I \times G_F$, the local symmetry
group becomes $Stab(\lambda_I) \times Stab (\lambda_F)$, and we
perform a path integral over both particle trajectories $g_I$ and
$g_F$. The conserved charges and equations of motion are easily
determined as in the previous section. Let's now introduce
an interaction term.
\subsection*{The rigid string tension}
Although we have been arguing continuously with the intuitive image of an open
string in the back of our minds, we have not yet demonstrated the precise
connection between our construction and the physics of open strings. We will
make the connection precise in this subsection.  The rigid string we construct
will correspond to two charged particles connected by a spring. To represent
the spring, we add a potential term to the action that is proportional to the
distance squared between the two endpoints of the string. We suppose that a
non-degenerate bi-invariant metric on the Lie algebra exists, and we measure
the distance between the end-points in this metric. The metric allows us to
identify the Lie algebra with its dual. The bracket operation $\langle . , .
\rangle$ then represents evaluation of the norm in the invariant
metric\footnote{In a matrix representation of a semi-simple algebra, the
  metric contraction is simply given by the trace: $\langle A,B \rangle =
  \text{Tr} (AB)$.}. The action is:
\begin{eqnarray}
S &=& \int_{L} d \tau \left( < \lambda_I, g_I^{-1} \partial_\tau g_I >
          -< \lambda_F, g_F^{-1} \partial_\tau g_F > \right)
\nonumber \\
& & 
     + \frac{K}{2} \int_{L} d \tau < g_I \lambda_I g_I^{-1}-g_F
     \lambda_F g_F^{-1}, g_I \lambda_I g_I^{-1}-g_F \lambda_F g_F^{-1}>
\label{rigidstringaction}
\end{eqnarray}
Firstly, we notice that $g_{I,F}$ can still
be separately gauge transformed, and they live in the
manifold $G/Stab(\lambda_{I,F})$, respectively. The global symmetry
group is broken however, by the interaction term, to a diagonal left
action on both group elements.\footnote{This is to be compared to the
  symmetry of a usual spring in flat space under overall translation.}

The classical dynamics of the system is solved for as follows. The
equations of motion are:
\begin{eqnarray}
\partial_\tau (g_I \lambda_I g_I^{-1}) + K {[} g_I \lambda_I g_I^{-1} ,
                                             g_F \lambda_F g_F^{-1} {]}
&=& 0 
\nonumber \\
-\partial_\tau (g_F \lambda_F g_F^{-1}) + K {[} g_F \lambda_F g_F^{-1} ,
                                             g_I \lambda_I g_I^{-1} {]}
&=& 0
\end{eqnarray}
where ${[} , {]}$ denotes the commutator. We can read off from these
equations the conserved currents associated to the global symmetry.
We
define the (non-conserved) charges $I_I=g_I \lambda _I g_I^{-1}$ and
$I_F = -g_F \lambda_F g_F^{-1}$ which generate the same algebra
(because of the opposite sign of the charges and of the symplectic
structures for the final point compared to the initial point). 
Note that the second charge $I_F$ corresponds to {\em minus}
 the position of the
final end-point of the string.
The sum
of these charges is conserved:
\begin{eqnarray}
\partial_\tau (I_I+I_F) &=& 0
\end{eqnarray}
It generates the simultaneous translation $g_{I,F} \mapsto h g_{I,F}$
where $h$ is any element of $G$. We reformulate the equations of
motion in terms of these charges:
\begin{eqnarray}
\partial_\tau I_I - K {[} I_I , I_F {]}
&=& 0 
\nonumber \\
\partial_\tau I_F + K {[} I_I , I_F {]}
&=& 0
\end{eqnarray}
We already know that we need to take $I_I+I_F$ constant to solve the
equations of motion. The difference $I_I-I_F$ can now be computed from
the equations of motion. We rewrite:
\begin{eqnarray}
\partial_\tau (I_I+I_F) &=& 0 \nonumber \\
\partial_\tau (I_I-I_F) &=& 2 K {[} I_I,I_F {]} 
                         = K {[} I_I-I_F, I_I + I_F {]}
\nonumber \\
I_I -I_F &=&   e^{-\tau K(I_I+I_F)}  (I_I-I_F)_0  e^{\tau K(I_I+I_F)}
\end{eqnarray}
The last equation gives the solution to the classical dynamics, given
a constant charge $I_I+I_F$, and an initial condition $(I_I-I_F)_0$.

The conserved quantity $I_I+I_F$ is interpreted as the length vector
of the string. Indeed, it measures the fixed difference
vector (in the Lie algebra) between the initial and final points of
the string. The motion is then dictated by conjugation of the initial
vector $(I_I-I_F)_0$ by the group element which is the exponential of
the length vector times the elapsed time times the parameter $K$.

\begin{figure}
\centering 
\includegraphics[width=5cm]{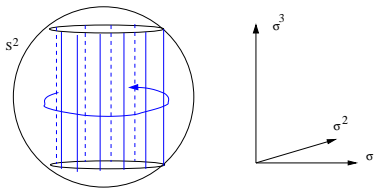}
\caption{A rigid string beginning and ending on a given $su(2)$ orbit,
  and rotating.}
\label{stringinmotion}
\end{figure}

Let us give one example to illustrate the concreteness of the above
solution (which is generically valid). Consider our favorite $SU(2)$
example, and consider for example a fixed orbit
$\lambda=\lambda_I=\lambda_F$ and a length vector $I_I+I_F$
proportional to $\sigma_3$. Conjugation of a vector proportional to
$\sigma_1$, say, will then lead to a velocity in the $\sigma_2$
direction. We get the following picture. Take a rigid string with its
endpoints on the sphere $\lambda$. It rotates around the central axis
parallel to this string (see figure~\ref{stringinmotion}). The
velocity is dictated by the string tension and its charge.  (One can
combine figures~\ref{magnmonandcharge} and \ref{stringinmotion}, and
the analogy between these motions and the motion of strings in infinite flat
branes with a $B-$field, discussed in detail in \cite{Bigatti:1999iz},
is clear. Namely, we have a positively and a negatively charged particle,
connected by a spring, in a constant magnetic field. Here however, we have a
curved phase space of finite volume.)

\begin{figure}
\centering 
\includegraphics[width=5cm]{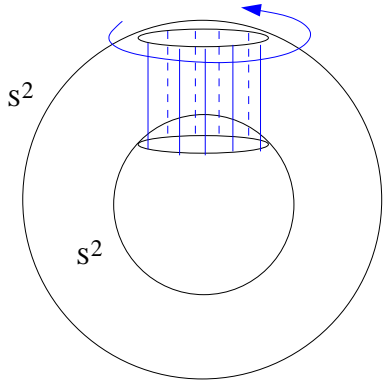}
\caption{A rigid string stretching between two different $su(2)$
  orbits, and rotating.}
\label{stringonorbits}
\end{figure}

The motion of the individual endpoints is:
\begin{eqnarray}
\Delta & \equiv & I_I+I_F 
\nonumber \\
I_I &=& \frac{1}{2} e^{-\tau K \Delta}  (I_I-I_F)_0  e^{\tau K \Delta}
+ \frac{\Delta}{2}
\nonumber \\
I_F &=& -\frac{1}{2} e^{-\tau K \Delta}  (I_I-I_F)_0  e^{\tau K \Delta}
+ \frac{\Delta}{2}
\label{motionendpoints}
\end{eqnarray}
We can also be more explicit about further constraints. Both $I_{I,F}$
should lie in a given sphere since $\langle I_{I} , I_{I} \rangle =
\langle \lambda_I, \lambda_I \rangle $, and similarly for the final
point.  This puts constraints on the relation between the initial
length vector and the initial velocity, which can be computed.
For example, for a fixed orbit $\lambda=\lambda_I=\lambda_F$ in
$SU(2)$ the constraints will say that the stretching of the string
(i.e. the relative coordinates of the initial and final points) must
be orthogonal to the initial center of mass location, and that the
size of the string is fixed in terms of its center of mass coordinate
(and the size of the fixed orbit $\lambda$). These constraints can
easily be interpreted in the case of $SU(2)$ from the figure
\ref{stringinmotion}. Another illustration of a motion in the case of
an open string ending on two different orbits is given in
figure~\ref{stringonorbits}.
\subsection*{A check}
We would like to check
directly that our construction agrees with a limit of 
the dynamics of open
strings on group manifolds as described in string theory. To that end
we briefly remind the reader of some aspects of strings on group
manifolds (see e.g. \cite{Alekseev:1998mc}). As for the bulk
Wess-Zumino-Witten model on group manifolds, the solutions to the
classical equations of motion for the open string embedding factorize
into a left-moving and a right-moving part:
\begin{eqnarray}
g &:& \text{strip} \ \mapsto G \nonumber \\
& & (\sigma, \tau) \mapsto g(\sigma, \tau) = g_+(\sigma +
\tau) g_-( \sigma- \tau)
\end{eqnarray}
where the $\text{strip}$ is the product of intervals ${[} -\frac{\pi}{2} ;
\frac{\pi}{2} {]} \times \mathbb{R}$. We can define left-moving and
right-moving conserved currents:
\begin{eqnarray}
J_+ &=& - \partial_+ g \ g^{-1}
\nonumber \\
J_- &=& g^{-1} \partial_- g
\end{eqnarray}
where we made use of the coordinates $ x^{\pm} = \sigma \pm \tau$.
The boundary conditions for open strings ending on conjugacy classes
\cite{Alekseev:1998mc} are that the left-moving and right-moving
current are related by $J_+=-J_-$ at the endpoints of the open string
at $\sigma=\pm \pi/2$. We wish to concentrate on solutions to the open
string equations of motion that are rigid (i.e. that have no
oscillatory excitations). We thus propose the ansatz:
\begin{eqnarray}
g_+ &=& e^{x^+ T_+} g_+^0
\nonumber \\
g_- &=& (g_-^0)^{-1} e^{x^- T_-}
\end{eqnarray}
The boundary conditions then enforce that $T_+=T_-=T$, where $T$
belongs to the Lie algebra. We thus find the solution to the equations
of motion with the boundary conditions (where $g^0=g_+^0
(g_-^0)^{-1}$):
\begin{eqnarray}
g(\sigma, \tau) = e^{x^+ T} g^0 e^{x^- T} = e^{\tau T} ( e^{\sigma T}
g^0 e^{\sigma T} ) e^{- \tau T}
\end{eqnarray}
The latter form of the solution shows that every rigid string bit (at
fixed value for $\sigma$) moves along its own conjugacy class. In the
rigid open string limit of our interest \cite{Alekseev:1999bs}, the
motion of the string is concentrated near a given point (say, the
identity) of the group manifold. We can implement this in the study of
our classical solution by assuming that $e^{\sigma T} g^0 e^{\sigma
  T}$ is near the identity. By putting $g^0=e^{X^0}$ and $g=e^X$, we
get -- note when performing the expansion near the identity that the
range of $\sigma$ is bounded, while that of $\tau$ is not --:
\begin{eqnarray}
X(\sigma,\tau)  &=& e^{\tau T} (X^0 + 2 \sigma T) e^{-\tau T}
\label{solutionopenstring}
\end{eqnarray}
It is clear that the two end-points of the open string now behave
precisely as in the system described above, namely, they move along
orbits in the Lie algebra in accord with the solutions to the
equations of motion for our two-particle system. One can identify the
parameters $T=K \Delta$, etc. in equations (\ref{motionendpoints})
and (\ref{solutionopenstring}). The comparison to the
previous section shows that the charges $I_{I,F}$ are indeed the
positions of the endpoints of the string on the Lie algebra. In
appendix \ref{rigidlimit} we give some further discussion of how the
rigid open string limit is taken in Wess-Zumino-Witten models
\cite{Alekseev:1999bs}.

We also recall here the results of the neat paper
\cite{Morariu:2004aw} (which is based on the intuition for rigid
open strings on spheres described in \cite{McGreevy:2000cw}). The
paper \cite{Morariu:2004aw} carefully defines and analyzes the rigid
open string limit (for the $SU(2)$ case only) starting from the
Wess-Zumino-Witten action for open strings. It shows how in the
scaling limit where the volume of the brane is kept fixed, while the
bulk is flattened (see appendix \ref{rigidlimit}), the action for the
open string reduces to an action for two oppositely charged point
particles interacting through a spring.  The action found in
\cite{Morariu:2004aw} after the limiting procedure coincides with the
one we have constructed in equation~(\ref{rigidstringaction}) when
restricted to the case of the group $G=SU(2)$. 
The one-form arises from the electromagnetic
field on the D-branes, while the Hamiltonian arises from the bulk
length term, and the kinetic term vanishes (as in flat space
\cite{Bigatti:1999iz}) in the
limit. Now, the verification we performed of the
solutions to the equations of motion is equivalent to the analysis of
\cite{Morariu:2004aw} in the $SU(2)$ case, and shows that the limiting
procedure on the action generalizes to all groups with a non-degenerate
invariant metric.

Thus, we have argued in detail from first principles that our system
does describe the classical open string dynamics in the rigid limit.
We can thus be confident that the quantization of the system also
faithfully represents aspects of quantum mechanical open string
theory.

We finally remark that the classical group dynamics that we
constructed can be generalized to other systems (not necessarily
descending from string theory). It would be interesting to analyze for
instance other interaction Hamiltonians (beyond the "spring
Hamiltonian") consistent with the gauge symmetries of the two-particle
system, and to set up the general evaluation of gauge invariant
observables (for instance in a path integral formalism). This
gives a natural physical context to the extension of the
Kirillov formalism from irreducible to tensor product representations.

\subsection*{The rigid open string Hilbert space}
Thus far we have mostly discussed the classical rigid open string. We
have already observed that the quantum Hilbert space will be a tensor
product of irreducible representations $\lambda_I \otimes
\bar{\lambda_F}$. Moreover, the interaction term (proportional to the
tension of the string) breaks the $G \times G$ global symmetry to the
diagonal subgroup. It is this symmetry breaking that prompts us to
decompose the tensor product Hilbert space into representations that
are irreducible under the (unbroken diagonal) subgroup. We thus
decompose: $\lambda_I \otimes \bar{\lambda_F} = \sum_{L} {C_{IF}}^{L}
\lambda_{L}$ and we interpret the weight $\lambda_L$ as being
associated to the length of the string (and therefore also to the
position of the center of mass of the string). Indeed, note that the
conserved quantities classifying the representations will include the
quadratic Casimir of the diagonal symmetry group (-- which is
associated to the conserved charge $\langle I_I+I_F,I_I+I_F \rangle$
in the classical dynamics --), which is nothing but the length of the
string squared.  (After quantization this quadratic Casimir is
proportional to the conformal weight associated to the vertex operator
of the open string state.)

Note that there is a slight change in perspective in the way in which
the open string Hilbert space arises here. Were we to quantize the
open string ($\sigma$-model) action directly, then the different
components of the open string Hilbert space would arise from {\em
  integrating} over all possible lengths $T/K$ of the rigid open string.
From the two-particle perspective that we developed, the same open
string Hilbert space arises as a {\em tensor product} of two single
particle Hilbert spaces. After decomposing the tensor product Hilbert
space into irreducible representations, we note that the resulting
Hilbert spaces coincide.

\begin{figure}
\centering 
\includegraphics[width=5cm]{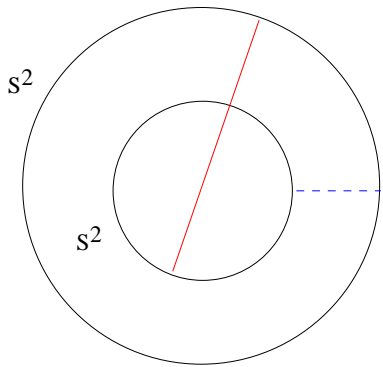}
\caption{A (blue dashed) string of minimal length and a (red) string of maximal
  length, stretching between two $su(2)$ orbits.}
\label{maxminlength}
\end{figure}
Returning to our favorite example, the case of $SU(2)$, and fixing for
instance that the open string begins and ends on one given D-brane labeled by
$j=j_I=j_F$, we find that we get a decomposition into representations of spins
$l=0,1,\dots,2j$ which correspond to a string of length zero up to the maximal
extension (root) $2j$.  The fact that only integer spins appear follows from
the fact that only half-integer spins can appear for a given endpoint, and
from the fact that we concentrated on a string that starts and ends on the
same brane $j$.  The quadratic Casimir will in this case be roughly
proportional to the length squared, as is the dimension of the (primary) open
string vertex operator~: $h_{open} = l(l+1)/(k+2)$.  In
figure~\ref{maxminlength} we illustrate a minimal and a maximal length string,
in the case of a string ending on two different orbits $j_I $ and $j_F$ of
$su(2)$, with length proportional to $|j_I-j_F|$ and $j_I+j_F$ respectively,
which correspond to the minimal and maximal spin occurring in the tensor
product decomposition $j_I \otimes j_F$.

Thus, to summarize, we have very general formulas for the
semi-classical open string spectrum between {\em any} two (symmetric) branes
(that correspond to conjugacy classes on group manifolds, or rather,
orbits in the Lie algebra), and we understand the quantization in
terms of symplectic geometry and geometric quantization.  The
generality of the formulas follows from our understanding of the
quantization of (co-)adjoint orbits.  A point that we retain from the
semi-classical picture is that the breaking of the product group to
the diagonal symmetry group by the open string Hamiltonian gives a
physical rationale for decomposing the tensor product representation
(since we tend to work in a basis where the Hamiltonian is diagonal).

\section{The interactions and a program}
\subsection{Interactions}
\label{interactions}
We have analyzed the free classical dynamics of open strings, and we
have quantized the free string. In the case where we have a
finite-dimensional Hilbert space, its dimension is the product of the
dimensions of the irreducible representations of which it is the tensor
product. (In general, we have a space of linear maps between
irreducible Hilbert spaces.)  We obtain a $d(\lambda_I) \times d
(\bar{\lambda_F})$ matrix representing each string state for an open
string stretching between orbits $\lambda_{I,F}$
 (where $d(\lambda_I),
d(\bar{\lambda_F})$ denote the dimensions of the representations
associated to the weights $\lambda_I, \bar{\lambda_F}$ respectively).

In this section, we turn to the interactions of open strings. Since
open strings interact by combining and splitting, which happens when
open string endpoints touch, it is natural to assume that open string
interactions are coded by the multiplication of the above matrices (or
more generally by the composition of linear maps). The final end of a
first (oriented) open string will interact with the initial end of a
second (oriented) open string.
\begin{figure}
\centering 
\includegraphics[width=5cm]{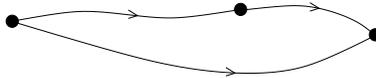}
\caption{Concatenating two strings produces a third.}
\label{concatenation}
\end{figure} This is the well-known picture that underlies
the intuition for open string field theory \cite{Witten:1985cc}.  See figure
\ref{concatenation}.

Since the open string interactions are communicated by their
endpoints, from the point of view of the center of mass of the open
strings, the interactions are non-local. Thus, when the interaction is
formulated in terms of the center of mass, we expect to have to take
into account an infinite number of derivative terms in coding the
interactions. This intuition is familiar from the flat space case
clearly explained in \cite{Bigatti:1999iz}.

Associativity of the open string interactions will be clear from the
associativity of matrix multiplication, or from the associativity of
the operation of concatenating open strings. However, when translated
in the non-local language adapted to the center of mass of the open
string, it is frequently
 less transparent. We will show later how it is
related to non-trivial identities in group theory and the theory of
special functions.
 
The associative product we thus construct is well-known for the case
of compact groups, from the analysis of boundary rational conformal
field theory, and from the associativity of the boundary vertex
operator algebra
\cite{Moore:1988qv,Pradisi:1996yd,Runkel:1998pm,Behrend:1999bn,Felder:1999ka,Felder:1999mq}.
For rational conformal field theories, which include
Wess-Zumino-Witten models on compact groups, the formulas below are
but a toy version of the boundary conformal field theory results.
Nevertheless, we will give the explicit description of the formalism,
and all precise factors and coefficients for the $SU(2)$ case, since
it is instructive, and since we had trouble localizing these simple
formulas in the literature in a consistent set of
conventions. Moreover, we will show later how this description
generalizes to other, non-rational cases that do not fall in the
framework of boundary (rational) conformal field theory as developed
hitherto.\footnote{We also provide in this way the semi-classical
  picture that was missing in the earliest days of "classical chiral
  conformal field theory" (where for instance, it was not entirely
  understood why the torus amplitude behaved badly in this limit --
  this is now explained by the fact that the limit pertains to the
  boundary conformal field theory, and that the bulk theory is
  completely flattened out and becomes of infinite volume --).}

\subsection{A mathematical program}
\label{program}
Before we turn to very concrete examples of our construction, we
briefly mention an even more general program that we can abstract from
the above considerations. We have mostly concentrated on co-adjoint
orbits that yield a symplectic foliation of Lie algebras \cite{Kirillov}.
However, quantization can be applied to much
more generic classes of manifolds. We can consider a given Poisson
manifold $L$, with four symplectic leaves
$M_{1,2,3,4}.$ 
Let's suppose that the symplectic form on these leaves is quantizable, and
that we can define pre-quantization bundles that admit a good polarization,
leading to Hilbert spaces $H_{1,2,3,4}$ corresponding to the manifolds
$M_{1,2,3,4}$. Then we can define operators ${O_i}^{j}$ which live in the
tensor product Hilbert space $H_i \otimes \bar{ H_j}$. The operator product is
defined as ${O_i}^j.{O_j}^k={O_i}^k$, where the product is determined by
tracing over the middle Hilbert space $H_j$.  The resulting operation is an
associative composition of operators associated to the four Hilbert spaces. It
would be interesting to analyze this product generically, and to establish its
connection with the analysis of symplectic groupoids.  Instead, we delve into
a few interesting examples.

\section{A compact example: $SU(2)$}
\label{su2}
In this section, we apply the formalism to the case of the compact Lie
group $SU(2)$ in great detail. We develop a diagrammatical language in
order to be able to show associativity of the resulting product in terms
of group theory.
\subsection*{The kinematical $3j$ symbols}
We have argued that we need to decompose the tensor product
representation into irreducible representations of the symmetry group.
This is a purely kinematical operation from the perspective of open
string theory.  In order to implement this decomposition, we can
choose a basis in the space of matrices (i.e. the tensor product
representation) that consists of the Clebsch-Gordan coefficients.
Since these are not symmetric under cyclic permutation, we prefer to
work with rescaled matrix elements, the $3j$ Wigner symbols. Due to
its symmetry properties, we can represent it by a (kinematic) cubic
vertex \footnote{Our diagrams are close cousins of those familiar from
  boundary conformal field theory, as we will review later. We also
  note here that closely related diagrams have been developed in
  \cite{Freidel:2001kb} (and references therein), in the context of
  spin networks. Our coding in diagrams will differ slightly from
  \cite{Freidel:2001kb}. We have chosen the correspondence
  between diagrams and $3j$ symbols to agree with the symmetries of
  the structure constants including the signs.  We will moreover
  take into account the metric on the space of $3j$ symbols as defined
  in \cite{Wigner}. The connection of these open string interactions
to spin
  networks deserves further study.}
\begin{eqnarray}
\lower0.2in\hbox{\input{3j1.pstex_t}}
=
\left( \begin{array}{ccc}
j_1 & j_2 & j \\
m_1 & m_2 & -m
\end{array} \right)
= \frac{(-1)^{j_1-j_2+m}}{\sqrt{2j+1}} C^{j_1 \ j_2 \ j}_{m_1 m_2 m}
\label{3jdef}
\end{eqnarray}
The explicit formula for the Clebsch-Gordan coefficients reads:
\begin{eqnarray}
C^{j_1 \ j_2 \ j}_{m_1 m_2 m} &=& (-1)^{j_1+j_2-j} \sqrt{2j+1}
\sqrt{\frac{(j_1-j_2+j)! (j_2-j_1+j)!}{(j_1+j_2+j+1)! (j_1+j_2-j)!}}
\nonumber \\
& & \times \sqrt{\frac{(j_1-m_1)! (j_2+m_2)! (j-m)! (j+m)!}{(j_1+m_1)!
    (j_2-m_2)!}} \\
& & \times \frac{1}{(j-j_1+m_2)! (j-j_2-m_1)!} \nonumber \\
& & \times {_3}F_2 \left( \begin{array}{c}
j-j_1-j_2, \ \ -j_1-m_1, \ \ -j_2+m_2 \\
j-j_1+m_2+1, \ \ j-j_2-m_1+1
\end{array} ;1 \right) \nonumber
\end{eqnarray}
In our conventions, $j_i$ and $m_i$ are half-integers, with $j_i$
positive and $-j_i~\leq~m_i~\leq~j_i$. Moreover, $m_1+m_2-m=0$ and
$|j_1-j_2| \leq j \leq j_1+j_2$. If these conditions are not
satisfied, the Clebsch-Gordan coefficient is zero\footnote{This result
  comes from the expression of the 3j symbol and the appearance of
  infinities in the $\Gamma$ functions when the above-stated
  conditions are not satisfied. The only exception is the condition
  that $m_1+m_2-m=0$, which should be enforced with a Kronecker
  symbol. We will however drop it for the sake of conciseness.}. Note
that conservation of the quantum number $m$ can be read very easily
from our diagrammatic notation. The dimension of the $SU(2)$
representation is $2j_i+1$ and the Casimir is $j_i (j_i+1)$. A useful
reference for Clebsch-Gordan and Racah coefficients (or equivalently
3j and 6j Wigner symbols) of the $SU(2)$ group is \cite{Vilenkin}.

One remark should be made concerning the arrows that appear in our
diagrammatic notation, since their direction does matter. More
precisely, changing the direction of an arrow amounts to multiplying
the vertex by a propagator which is the 1j Wigner symbol introduced in
\cite{Wigner}:
\begin{eqnarray}
\lower0.1in\hbox{\input{1j1.pstex_t}}
= (-1)^{j+m} \ \delta_{m,-m^{\prime}} = (-1)^{j-m^{\prime}} \
\delta_{m,-m^{\prime}}
\end{eqnarray}
Although simply a sign in the case of $SU(2)$, it can become more
complicated for other groups. For instance, we have:
\begin{eqnarray}
\lower0.2in\hbox{\input{3j4.pstex_t}}
&=& \ \left( \begin{array}{ccc}
j_1 & j_2 & m \\
m_1 & m_2 & j
\end{array} \right) \\
&=& \ \left( \begin{array}{ccc}
j_1 & j_2 & j \\
m_1 & m_2 & -m
\end{array} \right) \
\lower0.15in\hbox{\input{1j2.pstex_t}} \nonumber
\end{eqnarray}
while for changing the direction of the $j_1$ arrow, one would have to
multiply the 3j symbol by $(-1)^{j_1+m_1}$.

Another remark is that the overall 
inward direction we chose for arrows in
$(\ref{3jdef})$ was arbitrary:
\begin{eqnarray}
\lower0.2in\hbox{\input{3j1.pstex_t}}
\hspace{.2cm} = \
\lower0.2in\hbox{\input{3j2.pstex_t}}
\end{eqnarray}
i.e one may as well choose to represent the 3j symbol by a vertex with
all arrows pointing outwards.

We now recall, in our diagrammatic notation, several important results for 3j
symbols.  Our notation immediately reflects their invariance under circular
permutation of its indices. Beside cyclic permutation, two other useful
symmetry properties of the 3j symbols are:
\begin{eqnarray}
\label{3jchangem}
\lower0.2in\hbox{\input{3j1.pstex_t}}
&=& \ (-1)^{j_1+j_2+j} \ \
\lower0.2in\hbox{\input{3j3.pstex_t}} \\
&=& \ (-1)^{j_1+j_2+j} \ \
\lower0.2in\hbox{\input{3j5.pstex_t}} 
\end{eqnarray}
Note that permuting any two branches, whatever the direction of the
arrows, always amounts to a multiplication by $(-1)^{j_1+j_2+j}$,
while $(\ref{3jchangem})$ is only true when all arrows point inwards
or outwards, otherwise extra signs appear due to 1j symbols.

In the following we will concatenate 3j symbols -- concatenation is
possible when arrow directions are aligned and the labels $j_i, m_i$
coincide. 
Concatenation
implies that we sum over all internal half-integers $m_i$. The sums will
always be finite since only a finite number of 3j symbols are
non-zero.

The 3j symbols satisfy orthogonality and completeness
relations\footnote{For the reader's convenience, we recall here these
  relations in a more familiar form, i.e in terms of Clebsch-Gordan
  coefficients:
\begin{eqnarray}
\sum_{j,m \in \frac{1}{2} \mathbb{N}} C_{m_1 m_2 m}^{j_1 \ j_2 \ j}
C_{m'_1 m'_2 m}^{j_1 \ j_2 \ j} = \delta_{m_1,m'_1} \delta_{m_2 m'_2}
\end{eqnarray}
\begin{eqnarray}
\sum_{m_1,m_2 \in \frac{1}{2} \mathbb{N}} C_{m_1 m_2 m}^{j_1 \ j_2 \
  j} C_{m_1 m_2 m'}^{j_1 \ j_2 \ j'} = \delta_{j,j'} \delta_{m,m'}
\end{eqnarray}}:
\begin{eqnarray}
\lower0.4in\hbox{\input{normalization1.pstex_t}}
=
\lower0.5in\hbox{\input{normalization2.pstex_t}}
= \ \frac{(-1)^{j_1+j_2+j}}{2j+1} \ \delta_{j, j^{\prime}} \
\delta_{m, m^{\prime}}
\label{jmnormal}
\end{eqnarray}
and
\begin{eqnarray}
\sum_j \ (2j+1) \ \
\lower0.2in\hbox{\input{normalization3.pstex_t}}
\hspace{.8cm} = 
\delta_{m_1, m_1^{\prime}} \ \delta_{m_2, m_2^{\prime}}
\label{mmnormal}
\end{eqnarray}

\subsection*{The dynamical $6j$ symbols}
We now recall some useful results concerning the 6j Wigner symbol. It
can be expressed in terms of four Clebsch-Gordan coefficients:
\begin{align}
\label{6j43j}
& \left\{ \begin{array}{ccc}
j_1 & j_2 & j_{12} \\
j_3 & j & j_{23}
\end{array} \right\}
\ = \ \frac{(-1)^{j_1+j_2+j_3+j}}{\sqrt{(2j_{12}+1) (2j_{23}+1)} (2j+1)}
  \\
& \hspace{1.5cm} \times \sum_{m_1,m_2,m_3 \in \frac{1}{2} \mathbb{N}} C^{j_1 \ j_2 \
  j_{12}}_{m_1 m_2 m_{12}} \ C^{j_{12} \ j_3 \ j}_{m_{12} m_3 m} \ C^{j_2
  \ j_3 \ j_{23}}_{m_2 m_3 m_{23}} \ C^{j_1 \ j_{23} \ j}_{m_1 m_{23} m}
  \nonumber
\end{align}
In terms of 1j and 3j symbols, in our diagrammatic notation,
$(\ref{6j43j})$ is simply:
\begin{eqnarray}
\label{6jplanar}
\left\{ \begin{array}{ccc}
j_1 & j_2 & j_{12} \\
j_3 & j & j_{23}
\end{array} \right\}
\ = \
\lower0.4in\hbox{\input{6j1.pstex_t}}
\end{eqnarray}
The 6j symbol has a large group of symmetry, consisting of 144
elements and generated by the transformations:
\begin{eqnarray}
\label{6jsym}
(1) \ \left\{ \begin{array}{l}
j_1 \leftrightarrow j_2 \\
j_3 \leftrightarrow j
\end{array} \right.
\ \ &,& \hspace{1cm}
(2) \ \left\{ \begin{array}{l}
j_1 \leftrightarrow j_{12} \\
j_3 \leftrightarrow j_{23}
\end{array} \right.
\\
(3) \ \left\{ \begin{array}{l}
j_1 \leftrightarrow j_3 \\
j_2 \leftrightarrow j
\end{array} \right.
\ \ &,& \hspace{1cm}
(4) \ \left\{ \begin{array}{l}
j_{1,2,3} \rightarrow l-j_{1,2,3} \\
j \rightarrow l-j
\end{array} \right.
\nonumber
\end{eqnarray}
where $l = \frac{1}{2} (j_1+j_2+j_3+j)$. The 6j symbol is invariant
under any of these transformations, therefore under any permutation of
its columns. An interesting well-known remark is that this group of
symmetries includes the symmetries of a tetrahedron 
\cite{Wigner},
\begin{figure}
\centering 
\input{tetraedre2.pstex_t}
\label{tetrahedron}
\caption{The 6j symbol as a tetrahedron.}
\end{figure}
i.e one should see our planar representation of the 6j symbol
$(\ref{6jplanar})$ as a projection of the tetrahedron pictured in
$(\ref{tetrahedron})$ on the plane $BCD$ (with properly added
arrows\footnote{It may be worth noting that we intentionally did not draw
  arrows on the tetrahedron. Doing so would result in a complication
  since inverting the tetrahedron means exchanging left and right and
  therefore changing directions of arrows, i.e actually multiplying
  the 6j symbol by an extra sign:
\begin{eqnarray}
\label{26j}
\lower0.4in\hbox{\input{6j1.pstex_t}}
=(-1)^{2 j_1+2 j_3}
\lower0.4in\hbox{\input{6j2.pstex_t}}
\end{eqnarray}
Note that the diagram on the right-hand side of $(\ref{26j})$ has the
same symmetry properties of the 6j Wigner symbol, listed in
$(\ref{6jsym})$.}). From this representation, it is clear that the
result should be invariant of which tip of the tetrahedron is used in
order to project and obtain a planar diagram. To be more precise, the
symmetries of the tetrahedron generate transformations (1), (2) and
(3) (but not transformation (4)), with the following correspondence:
\begin{itemize}
\item transformation (1) $\longleftrightarrow$ invert and project from tip A
\item transformation (2) $\longleftrightarrow$ invert and project from tip B
\item transformation (3) $\longleftrightarrow$ project from tip C
\item transformation (1), then (2) $\longleftrightarrow$ project from
  tip D
\end{itemize}
where inverting the tetrahedron means that we take its mirror image with
respect to the plane opposite to the tip which we will project.

In the following we will need two important formulas. The first one is
the recoupling identity, which can be seen as a toy s-channel - t-channel
duality:
\begin{eqnarray}
\label{recouplingid}
\lower0.4in\hbox{\input{recoupling1.pstex_t}}
&=& \ \sum_{j_{12} \in \frac{1}{2} \mathbb{N}} (2j_{12}+1)
(-1)^{-j_1+j_2+j_3+j} \lower0.4in\hbox{\input{6j1.pstex_t}}
\nonumber \\
& & \hspace{.5cm} \lower0.3in\hbox{\input{recoupling2.pstex_t}}
\label{recoupling}
\end{eqnarray}
This identity originates from the definition of the $3j$ and $6j$ symbols as
transformations of bases.  It is worth noting that this identity together with
the orthogonality relation $(\ref{jmnormal})$ implies formula $(\ref{6j43j})$
for the $6j$ symbol.
We will also need the Biedenharn-Elliott identity (which follows from the
associativity of the tensor product of representations):
\begin{eqnarray}
\sum_{j_{23} \in \frac{1}{2} \mathbb{N}} (2j_{23}+1) (-1)^{J} &
\lower0.3in\hbox{\input{biedenharn1.pstex_t}} \
\lower0.3in\hbox{\input{biedenharn2.pstex_t}} \
\lower0.3in\hbox{\input{biedenharn3.pstex_t}} \nonumber \\
& = 
\lower0.3in\hbox{\input{biedenharn4.pstex_t}} \
\lower0.3in\hbox{\input{biedenharn5.pstex_t}}
\label{biedenharn}
\end{eqnarray}
where $J = j_1+j_2+j_3+j_4+j+j_{12}+j_{23}+j_{34}+j_{123}+j_{234}$.
One last useful formula can be obtained from the recoupling identity
$(\ref{recouplingid})$ and the orthogonality property
$(\ref{jmnormal})$ and reads:
\begin{eqnarray}
\lower0.3in\hbox{\input{6j3.pstex_t}}
= (-1)^{j_1-j_2+j_{12}+2j}
\lower0.3in\hbox{\input{6j4.pstex_t}}
\ \
\lower0.3in\hbox{\input{3j6.pstex_t}}
\label{3jinteraction}
\end{eqnarray}

\subsection*{Associativity}
We now code the interaction of open strings, which is the matrix
product, in terms of our diagrammatic notation. Multiplication of
matrices corresponds to concatenating diagrams, with the convention
that internal lines are summed over.
We then compute the cubic rigid string interaction
vertex\footnote{It codes the operator product expansion of the
  boundary vertex operator algebra.}, which is a dynamical quantity. We indicate the different
nature of the cubic vertex from the kinematical one by color-coding:
bold blue branches indicate centers of mass (spin $l_i$), while black
branches indicate branes (spin $j_i$).

A string stretching between two $SU(2)$ orbits labeled by their spins
$j_1$ and $j_2$ respectively, with its center of mass in the
representation $l$, will be represented by the following matrix
elements:
\begin{eqnarray}
\left[ \Theta^l_m \right]^{j_1 j_2}_{n_1 n_2} = 
\lower0.2in\hbox{\input{3j7.pstex_t}}
\end{eqnarray}
\begin{figure}
\centering 
\input{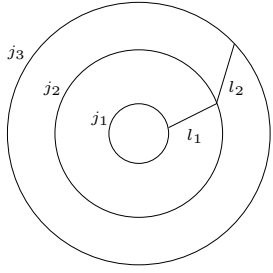}
\caption{Two strings interacting. \label{twostringsinteract}}
\end{figure}
Two strings may interact if their endpoints move in the same $SU(2)$
orbits, see figure~\ref{twostringsinteract}. This interaction
is encoded in the multiplication of the associated matrices, which can
be decomposed as a sum:
\begin{eqnarray}
\left[ \Theta^{l_1}_{m_1} \right]^{j_1 j_2}_{n_1 n_2} \times \left[
  \Theta^{l_2}_{m_2} \right]^{j_2 j_3}_{n_2 n_3} = \sum_{l_{12},
  m_{12} \in \frac{1}{2} \mathbb{N}} c_{l_{12}, m_{12}} \left[
  \Theta^{l_{12}}_{m_{12}} \right]^{j_1 j_3}_{n_1 n_3}
\end{eqnarray}
where $l_{12}$ is a representation in the tensor product of $l_1$ and
$l_2$, and where the coefficient $c_{l_{12}, m_{12}}$ can be computed
using $(\ref{3jinteraction})$ and $(\ref{jmnormal})$:
\begin{eqnarray}
c_{l_{12}, m_{12}} = (-1)^{\alpha_c} \ (2l_{12}+1) \
\lower0.3in\hbox{\input{6j5.pstex_t}}
\ \
\lower0.2in\hbox{\input{3j8.pstex_t}}
\label{productcoefficient}
\end{eqnarray}
where $\alpha_c = j_1+j_3+2l_2-l_{12}$ (See e.g.
\cite{Hoppe:1988gk,Alekseev:1999bs} for earlier
occurrences of this product for the case of a fixed orbit $j$).  Now
that we have the interaction vertex, and therefore the product of open
string operators, we can check associativity for the string
interaction directly in this non-local formalism (instead of using
 the map to the multiplication of matrices). The proof of the
associativity of the product reads:
\begin{align}
& \left( \left[ \Theta^{l_1}_{m_1} \right]^{j_1 j_2}_{n_1 n_2}
  \times \left[ \Theta^{l_2}_{m_2} \right]^{j_2 j_3}_{n_2 n_3} \right)
  \times \left[ \Theta^{l_3}_{m_3} \right]^{j_3 j_4}_{n_3 n_4}
\nonumber \\
&\hspace{2cm}= \sum_{l_{12}, m_{12}} c_{l_{12}, m_{12}} \left[
  \Theta^{l_{12}}_{m_{12}} \right]^{j_1 j_3}_{n_1 n_3} \times \left[
  \Theta^{l_3}_{m_3} \right]^{j_3 j_4}_{n_3 n_4}
\end{align}
\begin{align}
&= \sum_{l_{12}, l} \ (-1)^{2j_1+j_3+j_4+2l_2+2l_3-l_{12}-l}
(2l_{12}+1) (2l+1) \nonumber \\
& \hspace{1cm} \lower0.3in\hbox{\input{associativity1.pstex_t}} \
\lower0.3in\hbox{\input{associativity2.pstex_t}} \
\lower0.4in\hbox{\input{associativity3.pstex_t}}
\label{assoc1} \\
&=\sum_{l_{12}, l} \ (-1)^{2j_1+j_3+j_4-l_{12}-l_2-l_3+l_1-2l}
(2l_{12}+1) (2l_{23}+1) (2l+1) \nonumber \\
& \lower0.3in\hbox{\input{associativity1.pstex_t}} \
\lower0.3in\hbox{\input{associativity2.pstex_t}} \
\lower0.3in\hbox{\input{associativity7.pstex_t}} \
\lower0.2in\hbox{\input{associativity6.pstex_t}}
\label{assoc2} \\
&= \sum_{l_{23}, l} \ (-1)^{j_1+j_2+2j_4+2l_3-l_{23}-l} (2l_{23}+1)
(2l+1) \nonumber \\
& \hspace{.4cm} \lower0.3in\hbox{\input{associativity4.pstex_t}} \
\lower0.3in\hbox{\input{associativity5.pstex_t}} \
\lower0.2in\hbox{\input{associativity6.pstex_t}} \
\label{assoc3} \\
&= \sum_{l_{23}, m_{23}} c_{l_{23}, m_{23}} \left[
  \Theta^{l_1}_{m_1} \right]^{j_1 j_2}_{n_1 n_2} \times
\left[ \Theta^{l_{23}}_{m_{23}} \right]^{j_2 j_4}_{n_2 n_4} \nonumber \\
&= \left[ \Theta^{l_1}_{m_1} \right]^{j_1 j_2}_{n_1 n_2} \times
  \left( \left[ \Theta^{l_2}_{m_2} \right]^{j_2 j_3}_{m_2 m_3} \times
  \left[ \Theta^{l_3}_{m_3} \right]^{j_3 j_4}_{m_3 m_4} \right)
\end{align}
We have used the recoupling identity $(\ref{recoupling})$ to go from
$(\ref{assoc1})$ to $(\ref{assoc2})$ and the Biedenharn-Elliott
identity $(\ref{biedenharn})$ to go from $(\ref{assoc2})$ to
$(\ref{assoc3})$. The proof is very general. In particular, it
generalizes the diagrammatic 
proof of \cite{Freidel:2001kb} to the case of
differing initial and final orbits. It only makes use of generic
properties of groups like associativity of the tensor product
composition. We will thus be able to show that it applies
to cases not considered before in the literature.

\subsection*{The relation to (rational) boundary conformal field theory}
The diagrammar above can be viewed as a special case of the work of
\cite{Behrend:1999bn,Felder:1999ka} (and follow-ups) on the algebra of
boundary conformal field theories.  Our diagrams and diagrammatic
techniques are (for the case of $SU(2)$) semi-classical limits of the
analysis done in these papers for rational conformal field theories.

The fact that group theory forms a representation of classical chiral
conformal field theory data has been known since the seminal work
\cite{Moore:1988qv} on axioms of conformal field theory. (The 3j symbols
correspond to intertwiners and the 6j symbols to fusion matrices.)
However, it
has almost exclusively been applied to finite, or compact groups, and
to rational conformal field theories. A notable exception is the
treatment of the $H_4$ group in
\cite{D'Appollonio:2003dr,D'Appollonio:2004pm}.  Also, a physical
situation to which the classical limit applies was left undetermined
in \cite{Moore:1988qv} as well as in many other works on realizations
of chiral conformal field theory data. In particular, it was already
noticed in the work \cite{Moore:1988qv} that the classical limit of
chiral conformal field theory would not be applicable to bulk
quantities like the torus partition function. Presently, this is
understood from the fact that the chiral conformal field theory data
may be thought off as applying to open string dynamics, and that the
bulk theory flattens to a Lie algebra (of infinite volume).
 
We believe the simple model we developed gives a neat physical picture
of all the relevant ingredients, even illuminating the (well-known)
formulas of the compact case.  That this extra intuition is useful
will become clear in the following.  Indeed, our construction works
for any group, including non-compact groups.  This implies in
particular that we can treat the semi-classical limit of models that
fall {\em outside} the framework of rational conformal field theory
considered in \cite{Behrend:1999bn,Felder:1999ka} (see also
\cite{Moore:1988qv,Pradisi:1996yd,Runkel:1998pm,Felder:1999mq}).

Note that it is traditional to represent the matrices $\Theta_m^l$ as
matrix elements of the corresponding spherical functions. In this way,
one obtains an associative product of spherical functions
(see equation (\ref{productcoefficient})). If we
concentrate on a fixed brane and the open strings living on it, we
obtain the fuzzy sphere function algebra \cite{Hoppe:1988gk,Madore:1991bw}.
 This point is further
discussed in the appendix \ref{berezin}. Note that this construction
generalizes easily to a function algebra for spherical functions
associated to different orbits, and to the case of spherical functions
on generic orbits (or flag manifolds). All these function algebras are
associative and non-commutative.

We have already given explicit formulas for the compact case of $S^2$
branes associated to $SU(2)$. Let us now proceed to produce new
results for the rigid open string limit of non-compact branes
associated to $SL(2,\mathbb{R})$ in order the illustrate the utility
of our approach. We will first give a geometric description of the
orbits, of the tensor product and of the interactions, then we will
produce explicit results.

\section{Remarks on $SL(2,\mathbb{R})$ orbits and products}
\label{sl2rbis}
In this section we develop our intuition on the kind of string
interactions that we will describe in section~\ref{noncompact} when we
will analyze associative products associated to $SL(2,\mathbb{R})$
orbits
.
\subsection{Orbits and representations}
\begin{figure}
\centering 
\includegraphics[width=9cm]{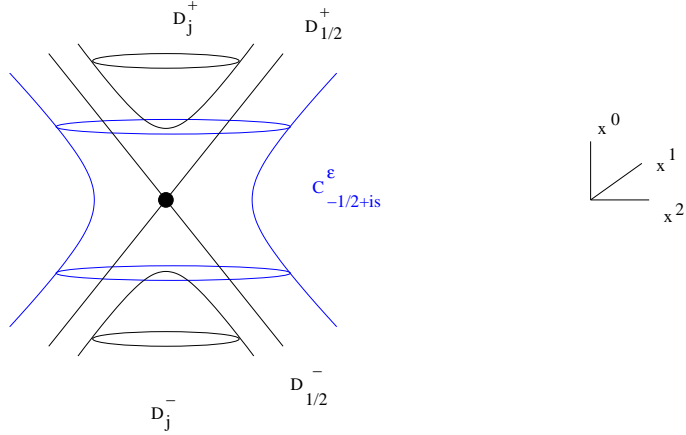}
\caption{The $sl(2,\mathbb{R})$ co-adjoint orbits. One-sheeted hyperboloids
  correspond to continuous representations and two-sheeted hyperboloids to
  discrete representations (the upper sheet being associated to positive
  discrete representations and the lower sheet to negative ones). The future
  and past light-cone (with the point at the origin removed) correspond to the
  special cases of discrete representations with $j=\frac{1}{2}$ while the
  point at the origin represents the trivial (identity) representation.}
\label{sl2rorbits}
\end{figure}

\begin{figure}
\centering 
\includegraphics[width=4cm]{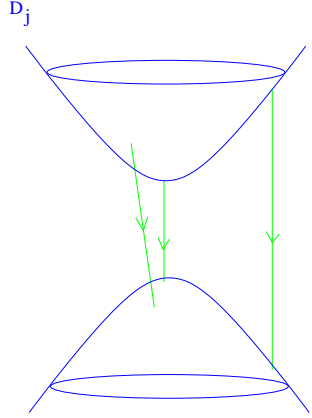}
\caption{Some past time-like strings starting on a future discrete orbit and
  ending on a past discrete orbit.}
\label{sl2rorbits3} 
\end{figure}

The co-adjoint orbits of $SL(2,\mathbb{R})$ are given in figure
\ref{sl2rorbits}. Vectors representing strings can be drawn between
these orbits (with their endpoints lying on the orbits), like in
figure \ref{sl2rorbits3} or \ref{sl2rorbits2} for instance.  We have
therefore future and past time-like and light-like vectors, and space-like
vectors. The time-like vectors correspond to discrete representations
while the continuous representations correspond to space-like vectors
(see e.g. \cite{Troost:2003ge} for a detailed discussion of the
correspondence).
Moreover, consider for instance a space-like oriented string which starts and
ends on a given (positive) discrete orbit (see figure \ref{sl2rorbits2}).
  Its first endpoint which is a
positively charged particle is then associated to a lowest weight
representation and its negatively charged second endpoint is associated to a
highest weight representation. Indeed, recall that we had to consider the
tensor product $\lambda_I \otimes \bar{\lambda}_F$.  There are strings
corresponding to any tensor product combination.

\subsection{The geometry of tensor product decomposition}
\begin{figure}
\centering 
\includegraphics[width=5cm]{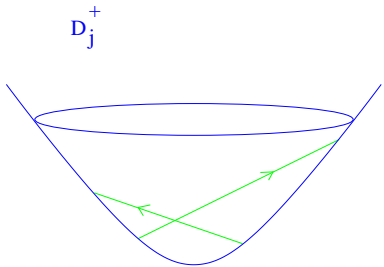}
\caption{A few space-like strings starting and ending on a given
  discrete orbit.}
\label{sl2rorbits2}
\end{figure}

To understand the irreducible representations in which the
center-of-mass wave-functions transform, we take a closer look at the
tensor product decompositions of $SL(2,\mathbb{R})$ representations
\cite{Mukunda:1974gb}:
\begin{eqnarray}
\label{decomposition}
\mathcal{D}^{\pm}_{j_1} \otimes \mathcal{D}^{\pm}_{j_2} &=& \oplus_{j
\geq j_1+j_2} \mathcal{D}^{\pm}_j
\\
\mathcal{D}^+_{j_1} \otimes \mathcal{D}^-_{j_2} &=& \int_0^{\infty} ds \
\mathcal{C}^{\epsilon}_s + \Theta (j_1-j_2-1)
\oplus_{j=j_{\text{min}}}^{j_1-j_2} \mathcal{D}^+_j
\nonumber \\
& & \hspace{1.5cm} + \Theta (j_2-j_1-1)
\oplus_{j=j_{\text{min}}}^{j_2-j_1} \mathcal{D}^-_j
\nonumber \\
& & \hspace{1cm} \text{where} \ \epsilon=0 ,\  j_{\text{min}}=1 \ \text{if} \ j_1+j_2 \
\text{is integer} \nonumber \\ 
& & \hspace{1cm} \text{and} \ \epsilon=\frac{1}{2} \ , \ j_{\text{min}}=\frac{3}{2}
\ \text{otherwise}
\nonumber
\end{eqnarray}
\begin{eqnarray}
\mathcal{D}^{\pm}_{j_1} \otimes \mathcal{C}^{\epsilon_2}_{s_2} &=&
\int_0^{\infty} ds \ \mathcal{C}^{\epsilon}_s \oplus_{j \geq
  j_{\text{min}}} \mathcal{D}_j^{\pm}
\nonumber \\
& & \hspace{1cm} \text{where} \ \epsilon=0 ,\  j_{\text{min}}=1 \ \text{if} \
j_1+\epsilon_2 \ \text{is integer} \nonumber \\ 
& & \hspace{1cm} \text{and} \ \epsilon=\frac{1}{2} \ , \ j_{\text{min}}=\frac{3}{2}
\ \text{otherwise}
\nonumber \\
\mathcal{C}^{\epsilon_1}_{s_1} \otimes \mathcal{C}^{\epsilon_2}_{s_2}
&=& \oplus_{j \geq j_{\text{min}}} \mathcal{D}_j^+ \oplus_{j \geq
  j_{\text{min}}} \mathcal{D}_j^- \oplus 2 \int_0^{\infty} ds \
\mathcal{C}^{\epsilon}_s
\nonumber \\
& & \hspace{1cm} \text{where} \ \epsilon=0 ,\  j_{\text{min}}=1 \ \text{if} \
\epsilon_1+\epsilon_2 \ \text{is integer} \nonumber \\ 
& & \hspace{1cm} \text{and} \ \epsilon=\frac{1}{2} \ , \ j_{\text{min}}=\frac{3}{2}
\ \text{otherwise}
\nonumber
\end{eqnarray}
where generically $j_i$ is a half-integer verifying $j_i \geq
\frac{1}{2}$, $\epsilon_i = 0$ or $\frac{1}{2}$ and $0 < s_i < \infty$.
The function $\Theta (x)$ is the Heaviside function that gives $1$ if $x \geq 0$
and $0$ otherwise. We have seen in the previous sections that we
should be able to interpret this tensor product decomposition
geometrically. Namely, the possible difference vectors (within the
vector space that is the Lie algebra) of the positions
of the open string stretching between two orbits are associated to
representatives of orbits that give rise to representations appearing
in the tensor product decomposition. Let us give an approximate
discussion of how this is consistent with the above list.
\begin{itemize}
\item If we take a  past time-like vector and a future time-like vector, then
  the difference is a past time-like vector. Its minimal length depends on
  the minimal lengths of these vectors linearly. That explains the first
  relation in (\ref{decomposition}). See figure~\ref{sl2rorbits3}.
\item The difference of two future time-like vectors can give either a
  space-like vector (see figure~\ref{sl2rorbits2}), or depending on
  their relative length, a future time-like or a past time-like
  vector. That geometry corresponds to the second relation in
  (\ref{decomposition}).
\item The difference of a future time-like vector and a space-like
  vector either gives a space-like vector, or a future time-like
  vector.  See the third relation in (\ref{decomposition}).
\item The difference of two space-like vectors can take any form,
as the tensor product in the last line in  (\ref{decomposition}).
\end{itemize}
This list is a rather sketchy explanation of the tensor product
decompositions, but we hope it is clear that the correspondence of the
representation theory and the geometric (quantization) picture can be
made precise.

\subsection{Interactions}
Since there are many sorts of (rigid open) strings we can realize
associative products arising from string concatenation in many
different spaces. In the following section we will concentrate on a
very particular case, for simplicity, although the construction is
generic for all discrete and continuous representations.

Let us consider the possibility where all the representations involved
are discrete representations. We can consider an open string
stretching from a (upward oriented) discrete 
orbit  to another (upward oriented) discrete
orbit, corresponding to a tensor product
representation $\mathcal{D}^+_{j_1} \otimes \mathcal{D}^-_{j_2}$. We
then consider a second string stretching from the 
$\mathcal{D}^+_{j_2}$ orbit to a third orbit
corresponding to a tensor product representation $\mathcal{D}^+_{j_2}
\otimes \mathcal{D}^-_{j_3}$. The resulting concatenated string will
be in the representation $\mathcal{D}^+_{j_1} \otimes
\mathcal{D}^-_{j_3}$. If we take $j_1 \geq j_2 \geq j_3$, the tensor
product decomposition for all three strings will contain
representations $\mathcal{D}^+_j$. That gives an intuitive picture of
the sort of concatenation that underlies the associative products we
will concentrate on, and their generalizations.

\section{A non-compact example: $SL(2,\mathbb{R})$}
\label{noncompact}
Note that $SL(2,\mathbb{R})$ is isomorphic to $SU(1,1)$. Its group
coefficients have been extensively studied in the literature
\cite{Liskova:1991xf,Davids:1998bp,Davids:2000kz}.  Our conventions
for discrete representations are that $j$ and $m$ are half-integers
with $m \geq j$ and $j, m \geq 0$.  Positive discrete series will be
labeled by $j$ and $m$ while negative discrete series will be
labeled by $j$ and $-m$. The study of $SU(1,1)$ is made easier by its
connection to $SU(2)$. For instance, 3j and 6j symbols of $SU(1,1)$
are related to 3j and 6j symbols of $SU(2)$ \cite{Liskova:1991xf}:
\begin{eqnarray}
\label{3jsu2su11}
\left[ \begin{array}{ccc}
j_1 & j_2 & j \\
m_1 & m_2 & -m
\end{array} \right]^{+++}_{SU(1,1)} =
\left( \begin{array}{ccc}
j_1^{\prime} & j_2^{\prime} & j^{\prime} \\
m_1^{\prime} & m_2^{\prime} & -m^{\prime}
\end{array} \right)_{SU(2)}
\end{eqnarray}
where:
\begin{align}
\label{identification3j}
&j_1^{\prime} = \frac{1}{2} \left( -j_1+j_2+m_1+m_2-1 \right)
&j_1 =& \frac{1}{2} \left(
  -j_1^{\prime}+j_2^{\prime}+m_1^{\prime}+m_2^{\prime} +1 \right)
\nonumber \\
&j_2^{\prime} = \frac{1}{2} \left( j_1-j_2+m_1+m_2-1 \right)
&j_2 =& \frac{1}{2} \left(
  j_1^{\prime}-j_2^{\prime}+m_1^{\prime}+m_2^{\prime} +1 \right)
\nonumber \\
&j^{\prime} = j-1
&j =& j^{\prime}+1 \\
&m_1^{\prime} = \frac{1}{2} \left( j_1+j_2-m_1+m_2-1 \right)
&m_1 =& \frac{1}{2} \left(
  j_1^{\prime}+j_2^{\prime}-m_1^{\prime}+m_2^{\prime} +1 \right)
\nonumber \\
&m_2^{\prime} = \frac{1}{2} \left( j_1+j_2+m_1-m_2-1 \right)
&m_2 =& \frac{1}{2} \left(
  j_1^{\prime}+j_2^{\prime}+m_1^{\prime}-m_2^{\prime} +1 \right)
\nonumber \\
&m^{\prime} = j_1+j_2-1
&m =& j_1^{\prime}+j_2^{\prime}+1
\nonumber
\end{align}
and:
\begin{eqnarray}
\left[ \begin{array}{ccc}
j_1 & j_2 & j_{12} \\
j_3 & j & j_{23}
\end{array} \right]^{+++}_{+++, \ SU(1,1)} =
\left\{ \begin{array}{ccc}
j_1^{\prime} & j_2^{\prime} & j_{12}^{\prime} \\
j_3^{\prime} & j^{\prime} & j_{23}^{\prime}
\end{array} \right\}_{SU(2)}
\end{eqnarray}
where:
\begin{align}
&j_1^{\prime} = \frac{1}{2} \left( j_1+j_2+j_3+j \right) -1
\hspace{1cm}
&j_1 =& \frac{1}{2} \left(
  j_1^{\prime}-j_2^{\prime}-j_3^{\prime}+j^{\prime} \right)
\nonumber \\
&j_2^{\prime} = \frac{1}{2} \left( -j_1-j_2-j_3+j \right)
\hspace{2cm}
&j_2 =& \frac{1}{2} \left(
  j_1^{\prime}-j_2^{\prime}+j_3^{\prime}-j^{\prime} \right)
\nonumber \\
&j_{12}^{\prime} = j_{12}-1
\hspace{4cm}
&j_{12} =& j_{12}^{\prime}+1
\\
&j_3^{\prime} = \frac{1}{2} \left( -j_1+j_2+j_3+j \right) -1
\hspace{1cm}
&j_3 =& \frac{1}{2} \left(
  -j_1^{\prime}-j_2^{\prime}+j_3^{\prime}+j^{\prime} \right)
\nonumber \\
&j^{\prime} = \frac{1}{2} \left( j_1-j_2+j_3+j \right) -1
\hspace{2cm}
&j =& \frac{1}{2} \left(
  j_1^{\prime}+j_2^{\prime}+j_3^{\prime}+j^{\prime} +1 \right)
\nonumber \\
&j_{23}^{\prime} = j_{23}-1
\hspace{4cm}
&j_{23} =& j_{23}^{\prime}+1
\nonumber
\end{align}
Note that the $+++$ signs that appear in our notation of the 3j and 6j
symbols indicate that all representations are in the positive discrete
representations. Negative discrete representations will be indicated
with a minus sign. In the following, we will only consider 3j and 6j
symbols of $SU(1,1)$, and will therefore drop the $SU(1,1)$ subscript.
It is also worth noting that the conditions for the 3j symbols to be
non-zero transform correctly from $SU(2)$ to $SU(1,1)$ through the
identification $(\ref{identification3j})$, for instance $j^ {\prime}
\geq m^ {\prime}$ corresponds to $j \geq j_1+j_2$. The explicit
formula for the 3j symbol is:
\begin{align}
\label{3jsu11}
&\left[ \begin{array}{ccc}
j_1 & j_2 & j \\
m_1 & m_2 & -m
\end{array} \right]^{+++} = (-1)^{2j_2-j+m-1} \sqrt{(j+j_1+j_2-2)!}
\\
& \hspace{2cm} \times \sqrt{(j+j_1-j_2-1)! (j-j_1+j_2-1)!
  (j-j_1-j_2)!}
\nonumber \\
& \hspace{2cm} \times \sqrt{\frac{(m_1+j_1-1)!
    (m_1-j_1)!}{(m_2+j_2-1)! (m_2-j_2)! (m+j-1)! (m-j)!}}
\nonumber \\
& \hspace{2cm} \times \frac{1}{(j+j_1-m_2-1)! (j-j_1-m_2)!}
\nonumber \\
& \hspace{2cm} \times \ _3F_2 \left(
\begin{array}{c}
j-m, \ \ 1-m_2-j_2, \ \ j_2-m_2 \\
j+j_1-m_2, \ \ j-j_1-m_2+1
\end{array} ; 1 \right) \nonumber
\end{align}
Because of the close connection between $SU(2)$ and $SU(1,1)$, all the results
that we used in the case of $SU(2)$ and that are needed for our discussion
remain valid in the case of $SU(1,1)$ \cite{Liskova:1991xf}. For instance, the
Biedenharn-Elliott identity for $SU(1,1)$ reads:
\begin{eqnarray}
&\sum_{l_{12}} (2l_{12}+1) \ (-1)^J
\left[ \begin{array}{ccc}
l_2 & l_1 & l_{12} \\
j_1 & j_3 & j_2
\end{array} \right]^{+++}_{---}
\left[ \begin{array}{ccc}
l_3 & l_{12} & l \\
j_1 & j_4 & j_3
\end{array} \right]^{+++}_{---}
\\
&
\times \left[ \begin{array}{ccc}
l_3 & l_2 & l_{23} \\
l_1 & l & l_{12}
\end{array} \right]^{+++}_{+++}
=
\left[ \begin{array}{ccc}
l_{23} & l_1 & l \\
j_1 & j_4 & j_2
\end{array} \right]^{+++}_{---}
\left[ \begin{array}{ccc}
l_3 & l_2 & l_{23} \\
j_2 & j_4 & j_3
\end{array} \right]^{+++}_{---}
\nonumber
\end{eqnarray}
where $J = j_1+j_2+j_3+j_4+l_1+l_2+l_3+l_{12}+l_{23}+l$. Note that one must
make sure that each spin is always in either a positive or a negative discrete
representation. The orthogonality, the completeness conditions and the
recoupling identity are also unchanged. Moreover, we have the following
equalities, valid up to a sign \cite{Davids:2000kz}:
\begin{eqnarray}
\label{transfo3jsu11}
\left[ \begin{array}{ccc}
j_1 & j_2 & j \\
m_1 & m_2 & -m
\end{array} \right]^{+++} &\sim&
\left[ \begin{array}{ccc}
j & j_1 & j_2 \\
-m & m_1 & m_2
\end{array} \right]^{-+-} \\
&\sim&
\left[ \begin{array}{ccc}
j_1 & j_2 & j \\
-m_1 & -m_2 & m
\end{array} \right]^{---}
\nonumber
\end{eqnarray}
together with:
\begin{eqnarray}
\label{symj1j2su11}
\left[ \begin{array}{ccc}
j_1 & j_2 & j \\
m_1 & m_2 & -m
\end{array} \right]^{+++} = (-1)^{m+j}
\left[ \begin{array}{ccc}
j_2 & j_1 & j \\
m_2 & m_1 & -m
\end{array} \right]^{+++}
\end{eqnarray}
We may then proceed and study the interaction.  Following our analysis
of $SU(2)$, the kinematics of strings is encoded in 3j symbols.
Because $SU(1,1)$ has several kinds of representations, we may
consider several distinct interactions, and use our geometric intuition
developed in the previous section. Any interaction which is consistent
with the tensor product decomposition $(\ref{decomposition})$ may be
considered. Note that the fact that the first endpoint of the string
is chosen to be in the representation of the brane on which it lives
while the second endpoint is in the conjugate representation is
essential for conservation of the quantum number $m$. In
this paper we will restrict ourselves to the explicit calculation of
the interaction involving discrete representations. More precisely, we
will focus on the study of:
\begin{eqnarray}
\left[ \Theta^l_m \right]^{j_1 j_2}_{n_1 n_2} = 
\left[ \begin{array}{ccc}
j_1 & j_2 & l \\
n_1 & -n_2 & -m
\end{array} \right]^{+-+} =
\lower0.2in\hbox{\input{3j7.pstex_t}}
\end{eqnarray}
Any other case is analogous to this one. A detailed look at
the computations shows that we do not need to know the signs involved
in the symmetry transformations of the 3j symbols
$(\ref{transfo3jsu11})$ in order to prove the associativity of the
product.\footnote{If needed these signs can be obtained from a
  detailed analysis of the Whipple relations for $_3F_2 (1)$
  hypergeometric functions. We have performed this analysis in appendix~\ref{whipple}.} Indeed, the
group structure ensures that:
\begin{eqnarray}
\left[ \Theta^{l_1}_{m_1} \right]^{j_1 j_2}_{n_1 n_2} \times \left[
  \Theta^{l_2}_{m_2} \right]^{j_2 j_3}_{n_2 n_3} = \sum_{l_{12},
  m_{12} \in \frac{1}{2} \mathbb{N}} c_{l_{12}, m_{12}} \left[
  \Theta^{l_{12}}_{m_{12}} \right]^{j_1 j_3}_{n_1 n_3}
\end{eqnarray}
and that the coefficient $ c_{l_{12}, m_{12}}$ satisfies:
\begin{eqnarray}
c_{l_{12}, m_{12}} \propto \lower0.2in\hbox{\input{3j8.pstex_t}}
\end{eqnarray}
Orthogonality and recoupling identities then ensure that, up to a
sign, the proportionality coefficient is precisely a 6j symbol 
(see (\ref{productcoefficient})). Using the above
relation, one may then see that all unknown signs cancel in
the proof of the associativity, therefore yielding:
\begin{align}
& \left( \left[ \Theta^{l_1}_{m_1} \right]^{j_1 j_2}_{n_1 n_2} \times \left[
  \Theta^{l_2}_{m_2} \right]^{j_2 j_3}_{n_2 n_3} \right) \times \left[
  \Theta^{l_3}_{m_3} \right]^{j_3 j_4}_{n_3 n_4} = \\
& \hspace{4cm} \left[ \Theta^{l_1}_{m_1} \right]^{j_1 j_2}_{n_1 n_2}
  \times \left( \left[ \Theta^{l_2}_{m_2} \right]^{j_2 j_3}_{n_2 n_3}
  \times \left[ \Theta^{l_3}_{m_3} \right]^{j_3 j_4}_{n_3 n_4} \right)
  \nonumber
\end{align}
We have therefore extended our analysis to a non-compact case,
and have given formulas for associative products for open string
wave-functions transforming in the discrete representations of
$SL(2,\mathbb{R})$. We have thus constructed fuzzy hyperboloids.

For future purposes we note
that all our computations extend to the quantum group
$U_q(SU(1,1))$. Recall for instance that in this case the 3j symbol
is a direct extension of the classical case \cite{Liskova:1991xf}:
\begin{align}
& \left[
\begin{array}{ccc}
j_1 & j_2 & j \\
m_1 & m_2 & m
\end{array}
\right]_q = (-1)^{j_1-m_1}
q^{\frac{1}{4} (j(j-1)+j_1(j_1-1)-j_2(j_2-1))-\frac{m_1(m-1)}{2}}
\nonumber \\
& \times \sqrt{[j-j_1-j_2]![j-j_1+j_2-1]![j+j_1-j_2-1]![j+j_1+j_2-2]!}
\nonumber \\
& \times \sqrt{
\frac{[m-j]![m_1-j_1]![m_1+j_1-1]![m_2-j_2]![m_2+j_2-1]!}{[m+j-1]!}}
\nonumber \\
& \sum_{n \geq 0} (-1)^n q^{\frac{n}{2} (m+j-1)}
\frac{1}{[n]![m-j-n]![m_1-j_1-n]![m_1+j_1-n-1]!} \nonumber \\
& \hspace{3cm} \cdot \frac{1}{[j-j_2-m_1+n]![j+j_2-m_1+n-1]!}
\end{align}
where the sum is taken over all $n$ such that integers in the sum
(say, $m-j-n$) are all positive, and we define:
\begin{eqnarray}
[n] =
\frac{q^{\frac{n}{2}}-q^{-\frac{n}{2}}}{q^{\frac{1}{2}}-q^{-\frac{1}{2}}}
\end{eqnarray}
This quantity is such that $[n] \rightarrow n$ when $q \rightarrow 1$.
It is then easy to see that this formula reduces to equation
$(\ref{3jsu11})$
in the classical limit $q \rightarrow 1$. In the same way, one can construct the 
 $U_q(SU(1,1))$ 6j-symbol \cite{Liskova:1991xf}. That defines again 
via equation (\ref{productcoefficient}) an associative product.

\section{Associative products based on quantum groups}
\label{quantumgroups}
In this section, we argue that our construction of associative
products related to rigid open strings also applies to the case where
we replace groups by quantum groups.

The construction of the associative product follows by now familiar
paths. However, we need the point-particle action replacing the action
for a particle on a co-adjoint orbit. The relevant symplectic form
(which can be integrated to the point-particle one-form Lagrangian) is
the one constructed by \cite{Alekseev:1993qs} in full generality (and
see earlier work \cite{Gawedzki:1990jc,Falcet:1991xt}), for any
Lie bi-algebra. It can be expressed in terms of Maurer-Cartan forms
associated to a algebra known as the Heisenberg double of the Lie
bi-algebra. (For more details see \cite{Alekseev:1993qs} and for a very
concrete example, we refer to \cite{Morariu:1997vu}.) 
Quantizing the Alekseev-Malkin action will provide for a Hilbert space
on which a natural set of observables  \cite{Morariu:1997vu}
acts irreducibly as quantum group generators.

By considering the sum of two such actions, for two independent
particles, one creates a quantum system with a Hilbert space which is
the tensor product of two irreducible representations of the quantum
group. An (open string, quadratic Casimir) Hamiltonian will break the
symmetry to a diagonal quantum subgroup, and the tensor product can be
decomposed accordingly.  More importantly, the tensor product
structure of the Hilbert space will naturally allow for defining an
associative product for wave-functions living within the tensor
product Hilbert space.  The associative product can be expressed in
terms of 3j and 6j symbols for the quantum group, exactly as was done
previously for ordinary Lie groups.  That is the construction which
 applies to at least all cases in which the symplectic form is
known \cite{Alekseev:1993qs}.

Furthermore, we claim that in the case where the quantum group is the
double of an ordinary Lie group allowing for a Wess-Zumino-Witten
model, the associative product we constructed coincides with the
associative product between primary boundary vertex operators living on
symmetry preserving branes.

Let us now collect the evidence for the above picture scattered over the literature:
\begin{itemize}
\item it is known that the symplectic form of \cite{Alekseev:1993qs}
  reduces to the Kirillov symplectic form in the classical limit thus
  allowing us to recuperate the rigid open string limit.  The
  symplectic form also has the required quantum group symmetry.
\item we can prove our claim in the case of $SU(2)$ by using the
  literature as stepping stones.
\begin{itemize}
\item{1.}  In the $SU(2)$ Poisson-Lie group case, it has been explicitly shown
  using a canonical analysis \cite{Gawedzki:1990jc,Falcet:1991xt,Chu:1994hm}
  that the symplectic form of \cite{Alekseev:1993qs} after quantization
  consistent with the symmetries gives rise to a Hilbert space that represents
  the $U_q(SU(2))$ quantum group irreducibly. Alternatively, a path integral
  analysis performed in detail in \cite{Morariu:1997vu} identifies the
  observables that act as a quantum group generators on the Hilbert space.
\item{2.}  Quantizing two such particle actions independently will
  give rise to a tensor product of irreducible representations of the
  quantum group.
\item{3.}  Following the procedure of this paper, we can show that the
  operators living in the tensor products compose according to the law
  governed by the 3j and 6j-symbols. The composition is associative since our
  proof in section \ref{su2} can be extended to the case of the quantum group
  $U_q(SU(2))$. Indeed, all the group identities that we used remain unchanged,
  up to minor substitutions.  (for instance, any number $n$ has to be replaced
  by $[n]$).
 %
All needed formulas can be found in
  \cite{Hou:1989jv,Hou:1989jt,Vilenkin3}.
\item{4.} The multiplication law coincides, we note a posteriori, with the
  multiplication law for boundary vertex operators, independently derived
  through entirely different methods in e.g. \cite{Alekseev:1999bs}.
(See also \cite{Podles:1987wd}.)
\end{itemize}
\item This should hold generically, at
  least for compact groups. 
Some analysis of dressing orbit quantization for generic
  quantum groups is in the mathematical literature (see e.g.
  \cite{JurcoStov1,JurcoStov2}).
\item A posteriori, it is clear from the solution to the
  Cardy-Lewellen constraints for compact WZW-models (see e.g.
  \cite{Felder:1999ka}), as well as for a particular non-compact
  WZW-model \cite{D'Appollonio:2003dr,D'Appollonio:2004pm} for
  symmetry preserving branes that the above products do coincide,
  since they have identical coefficients (given in terms of quantum 3j
  and 6j symbols). Thus, there is a proof of the above statement on a
  case-by-case basis, and as a heuristic guideline this is known in
  the boundary conformal field theory literature.
\end{itemize}

We believe we have added considerably to the understanding of the logic behind
the above equivalence. Indeed, in the large level limit, or, in other words,
near the group identity, the intuitive picture we developed proves the
equivalence.  Moreover, we believe that the fact that the deformed symplectic
form of \cite{Alekseev:1993qs} with the required quantum group symmetry exists
is evidence that our intuitive picture extends to the case of the quantum
group.  Note that we can thus code the dynamics of the full fluctuating open
string, in the (local) dynamics of two endpoints. Firstly we quantize the
endpoints in a way consistent with the quantum group symmetry,
determining the product of primary boundary vertex operators. Secondly, we use
the affine Kac-Moody symmetry to derive the operator product of descendents.

It is clear that the above statements lead to many new solutions to
the Cardy-Lewellen constraints (as for instance for symmetry preserving
branes in extended Heisenberg groups $H_{2n+2}$ \cite{Sfetsos:1994vz,Figueroa-O'Farrill:1994yf}), even in cases where one has
trouble defining the conformal field theory directly from an action principle.
A case in point is the $SL(2,\mathbb{R})$ conformal field theory,
which is difficult to define directly due to the Lorentzian signature
of the curved group. We refer to \cite{Maldacena:2001km,Ribault:2005ms}
for steps towards defining the theory via analytic continuation from the $H_3^+$
conformal field theory, or via modified Knizhnik-Zamolodchikov equations.
It is crucial to observe that our analysis in sections 
\ref{sl2rbis} and \ref{noncompact}
remains valid in the case of the
quantum group $U_q(SU(1,1))$, since all needed group identities remain
intact.
Thus, we construct a new solution to the
Cardy-Lewellen constraint that a future Lorentzian analytic
continuation of a $SL(2,\mathbb{C})/SU(2)$ boundary conformal field
theory will presumably need to match. This squares
well with the fact that the $6j$ symbols of the quantum group
form the basis of the solution for the boundary three-point function
of Liouville theory \cite{Ponsot:2000mt,Ponsot:2001ng},
when combined with the observation that in the bulk, Liouville theory and 
the $H_3^+$ model are already known to be closely related (see
e.g. \cite{Ribault:2005wp}).\footnote{Note that bosonic Liouville theory is
obtained from $SL(2,\mathbb{R})$ by gauging a light-like direction.}
(Recently it was shown in
 \cite{Hosomichi:2006pz} that the boundary correlation functions of 
the $SL(2,\mathbb{C})/SU(2)$ model can be computed in terms of bosonic
Liouville disc correlators.)

\section{Conclusions and open problems}
In this paper, we have added connections between subjects
that have an extended literature by themselves. In particular, we have
reviewed the connection of the orbit method in representation theory
to the quantization of a particle on an orbit, and its relation to
geometric quantization. Secondly, we  observed that we can apply this
construction to the two endpoints of an open string, and that we thus
obtain a tensor product of representations via the orbit method and
geometric quantization. Thirdly we noted that string concatenation leads
to an associative product for operators (or the associated functions)
living in the tensor product Hilbert spaces. The construction
has the considerable generality of the orbit method. 

We applied our formalism to the
known example of $SU(2)$, and we constructed
a product for the case of discrete representations of
$SL(2,\mathbb{R})$.
Also, in the case of $SU(2)$, we made more explicit the fact that the
fuzzy sphere is an example of Berezin quantization, which was left as
an exercise in the literature (see appendix~\ref{berezin}). We then
continued to make the connection to a formal star product, explicitly
verifying the existence of this limit, following the mathematics
literature on geometric quantization and star products.

Moreover, we argued that our construction extends to full solutions of boundary
conformal field theory, in particular including the case of non-compact
groups. We gave explicit new formulas for an associative product between
functions on a $U_q(SL(2,\mathbb{R}))$ quantum group orbit (which can now be verified as
a limiting case once the boundary conformal field theory is carefully defined and
solved).

We may summarize some avenues that cross different fields and that
might be useful to explore further:
\begin{itemize}
\item The intuitive picture of a string stretching between two
  co-adjoint orbits of a Lie group corresponding to an intertwiner
  between three representations (after tensor product decomposition)
  is attractive. In particular, it is natural to ask about an
  associative string interaction. For instance, we could apply this to
  intertwiners for representations of the Virasoro or Kac-Moody
  algebras, and ask for the meaning of the associative product of
  Virasoro or affine Kac-Moody intertwiners obtained in this way. That
  is now a natural mathematical question (beyond the scope of
  finite-dimensional target spaces for string theory).
\item We could further explore the geometry of D-branes as dressing
  orbits of quantum groups. This can be attacked by more directly linking
chiral conformal field theory to the theory of D-brane boundary states,
and in particular in regard to the quantum group symmetry (see e.g. 
\cite{Alvarez-Gaume:1988vr,Gawedzki:1990jc,Alekseev:1993nk,Falcet:1991xt} and
references therein). 
\item One would like to understand better the relation between the
  semi-classical approximation to the three-open string vertex
  operator product via a simple disc embedding into a given target
  space with D-branes, and the idea of formulating star products in
  terms of symplectic areas \cite{Weinstein}.
\item The program proposed in subsection~\ref{program}
  can be executed, and the precise connection to symplectic groupoids
 \cite{Weinstein96, Karasev86, Coste87, Weinstein90, Zakrzewski90}
  investigated.
\item One would like to investigate more the link between the
  associative products on $SL(2,\mathbb{R})$ orbits constructed here, and the associative
  products on the Poincar\'e disc constructed in the mathematics literature
\cite{Unterberger83, Unterberger88}. As formal power series of derivative
  operators acting on spherical functions, they are equivalent (and even have
equivalent star product cochains up to a constant), but one would like to
  understand better the relation when realized non-formally (within a
  specified function algebra, as in our construction).
\item We can repeat our construction for  twisted (co-)adjoint orbits
  \cite{Alekseev:2002rj}. 
\end{itemize}
In summary, we hope that our discussion of the quantization of pairs of
conjugacy classes (in particular of non-compact groups)
from various perspectives, including the string theoretic,
symplectic geometric and boundary conformal field theory viewpoint can lead
to further useful cross-fertilization.

\section*{Acknowledgements}
We would like to thank Costas Bachas and Kirill Krasnov for useful
discussions.
The work of J.T. is partially supported by the RTN European Programme
MRTN-CT2004-005104.

\appendix


\section{The rigid limit}
\label{rigidlimit}
The limit of string theory giving rise to the models developed
in section~\ref{pairs} of
 the bulk of the paper is the following \cite{Alekseev:1999bs}. We concentrate on
group manifolds with non-degenerate metric, and identify the Lie
algebra with its dual. We rescale the metric such that the group
manifold becomes flat. This involves sending the level of
the associated Wess-Zumino-Witten model to infinity.

However, at the same time, we keep the volume of the conjugacy class
under study fixed. This means it will be a conjugacy class very close
to the identity (say) of the group manifold. Near the identity, we can
approximate the group manifold by its tangent space, which is the Lie
algebra. The conjugacy class of an element near the identity is
equivalently described by the orbit of the associated element of the
Lie algebra. This is how the relevance of orbits in this limit becomes
manifest.

We note that the limit involved sends the NS-NS B-field and its
associated field strength to infinity, since these are proportional to
the level. 

\begin{figure}
\centering 
\includegraphics[width=10cm]{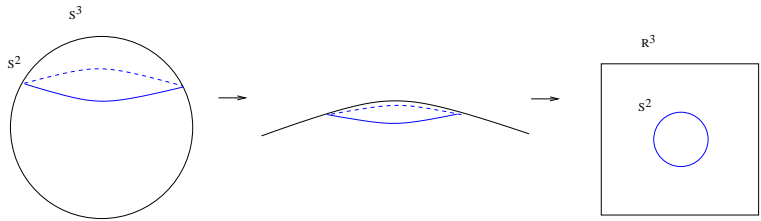}
\caption{The limit turning a conjugacy class into an orbit for the example
of $G=SU(2)$. We keep the size of the conjugacy class (orbit) fixed, while
rescaling the group metric until it becomes flat.}
\label{conjclassorbit}
\end{figure}

For the case of the group manifold $SU(2)$, this limit is visualized in figure
\ref{conjclassorbit}.
The open string oscillations decouple in this limit,
leaving us with a finite algebra of vertex operators corresponding to a rigid
open string \cite{Alekseev:1999bs}.

\section{Berezin quantization, coherent states and star products}
\label{berezin}
\subsection*{The relation to Berezin quantization}
We show that the algebra of matrices and spherical functions obtained
by the three-string interaction in section \ref{su2}  agrees with Berezin
quantization \cite{Berezin:1974} of the two-sphere for $SU(2)$. This shows that the
construction known as the fuzzy two-sphere
\cite{Hoppe:1988gk,Madore:1991bw} is nothing but Berezin's
quantization of a K\"ahler two-manifold, discussed in terms of
coherent states for instance in \cite{Perelomov:1986tf}.  We believe
this is perhaps known to experts (see e.g.
\cite{Hoppe:1988gk,Grosse:1993uq} for suggestions to that effect),
although it has not been stressed much, nor have we been able to find
the detailed map in the literature. Let us proceed to prove this
equivalence.

In the following, we first work in the formalism of
\cite{Perelomov:1986tf}, chapters 4 and 16 (mostly), and we will
assume familiarity with that reference.  Our strategy is to link
the Berezin quantization \cite{Perelomov:1986tf} to a quantization
discussed in reference \cite{Grosse:1993uq}, which is demonstrated there
to be equivalent to the fuzzy sphere. So, let's firstly perform Berezin
quantization of the sphere coset manifold $SU(2)/U(1)$ explicitly.  We
perform a stereographic projection and parameterize the sphere by a
complex variable $z$. The measure on the manifold is
\begin{eqnarray}
\label{zmeasure}
d \mu (z,\bar{z}) &=& \frac{2j+1}{\pi} (1+z \bar{z})^{-2j-2} dz  d
\bar{z}
\end{eqnarray}
Let's consider polynomial holomorphic functions of degree strictly
less than or equal to the positive integer $2j$, with the scalar
product:
\begin{eqnarray}
\langle f,g \rangle &=& \int_{\mathbb{C}} f(\bar{z}) g(z) d \mu (z, \bar{z})
\end{eqnarray}
One checks that the functions
\begin{eqnarray}
f_n (z) = \langle z | j,n \rangle = \sqrt{\frac{(2j)!}{n! (2j-n)!}} z^n 
\end{eqnarray}
for $n=0,1, \dots, 2j$ form an orthonormal basis of the Hilbert space
of dimension $2j+1$. They are in one-to-one correspondance with the
basis vectors $| j,n \rangle$ which satisfy:
\begin{eqnarray}
\langle j,n | j,n' \rangle = \langle f_n,f_{n'} \rangle &=&
\delta_{n,n'} \nonumber \\
\sum_{n=0}^{2j} | j,n \rangle \langle j,n | &=& 1
\end{eqnarray}
We define the coherent states:
\begin{eqnarray}
| \zeta \rangle = \sum_{n=0}^{2j} f_n (\zeta) (1+\zeta
  \bar{\zeta})^{-j} | j,n \rangle
\end{eqnarray}
where $\zeta \in \mathbb{C}$. They satisfy:
\begin{eqnarray}
\int_{\mathbb{C}} | \zeta \rangle \langle \zeta | (1+\zeta
\bar{\zeta})^{2j} d \mu (\zeta,\bar{\zeta}) = 1
\end{eqnarray}
where the measure is the one given in $(\ref{zmeasure})$. Note 
that:
\begin{eqnarray}
\int_{\mathbb{C}} | z \rangle \langle z | d \mu (z,\bar{z}) = 1
\end{eqnarray}
We also define the kernel $L$:
\begin{eqnarray}
L(\zeta,z) &=& \sum_{n=0}^{2j} f_n(z) f_n (\zeta) = (1+z \zeta)^{2j}
\nonumber \\
\langle  z | \zeta \rangle &=& L(z,\zeta) (1+\zeta \bar{\zeta})^{-j}
\end{eqnarray}
Using this kernel, it is possible to go from the $| z \rangle$ basis
to the $| \zeta \rangle$ basis and vice-versa (though both $z$ and
$\zeta$ are complex numbers, one should not confuse the associated
basis vectors that are different). For instance, $\langle \zeta | j,n
\rangle (1+\zeta \bar{\zeta})^j = f(\bar{\zeta})$. From the formula
for the kernel, we easily compute the norm:
\begin{eqnarray}
\langle \zeta' | \zeta \rangle = L(\zeta,\bar{\zeta}') (1+ \zeta
\bar{\zeta})^{-2j}
\end{eqnarray}
e.g. $\langle \zeta | \zeta \rangle = 1$.  Defining a symbol
associated to an operator $F$ in the Hilbert space, following Berezin:
\begin{eqnarray}
F(z, \bar{z}) &=& \frac{\langle \bar{z} | F | \bar{z}
  \rangle}{\langle \bar{z} | \bar{z} \rangle}
\end{eqnarray}
we can compute the symbol associated to the product of operators:
\begin{eqnarray}
F_{1.2} (z, \bar{z}) &=&
 \frac{\langle \bar{z} | F_1 F_2 | \bar{z} \rangle}{\langle \bar{z}  | \bar{z}
 \rangle} \\
&=& \int_{\mathbb{C}} \langle \bar{z} | F_1 | \bar{\zeta} \rangle  
\langle \bar{\zeta} | F_2 | \bar{z} \rangle \ \frac{(1+\zeta
 \bar{\zeta})^{2j}}{\langle \bar{z} | \bar{z} \rangle} d
 \mu(\zeta,\bar{\zeta}) \nonumber \\
&=& \int_{\mathbb{C}} F_1 (z, \bar{\zeta}) F_2 (\zeta,\bar{z})
 \frac{\langle \bar{z}| \bar{\zeta} \rangle \langle \bar{\zeta} |
 \bar{z} \rangle }{\langle \bar{z} | \bar{z} \rangle}
 (1+\zeta \bar{\zeta})^{2j} d \mu(\zeta,\bar{\zeta}) \nonumber \\
&=& \int_{\mathbb{C}} F_1 (z, \bar{\zeta}) F_2 (\zeta,\bar{z}) \left(
 \frac{ (1+z \zeta) (1+\bar{z} \bar{\zeta})}{(1+z \bar{z})}
 \right)^{2j} d \mu(\zeta,\bar{\zeta}) \nonumber
\end{eqnarray}
At this point, note that this definition of the product of
symbols (i.e. functions) associated to the product of (bounded)
operators in a given finite dimensional Hilbert space is the same as
the definition adopted by the authors of reference \cite{Grosse:1993uq} (see their
formula (8)). Thus, the product studied in \cite{Grosse:1993uq} {\em is} the
product of symbols in Berezin quantization, within the formalism of
coherent states. This is our first main point.

For comparison with various references in the literature, it can be useful to
express the product differently, using
a particular fixed  spin $j$ unitary irreducible representation of the group $SU(2)$, $T(g)$.
Using this representation, we can express operators $F$
in terms of functions $\tilde{f}$ on the group via the relation:
\begin{eqnarray}
F &=& \int dg \langle \bar{z} | T(g) | \bar{z} \rangle \tilde{f}(g)
\end{eqnarray}
We can then express the product of the corresponding symbols 
in terms of the (tilded) functions, if we so desire.

Also, given the representation $T(g)$ of the group, we can associate to it a
kernel function $\omega (g,\bar{z}) = \langle \bar{z} | T(g) | \bar{z}
\rangle / \langle \bar{z} | \bar{z}
\rangle$
which is the symbol of $T(g)$, for each $g$.  
The symbol $F(z, \bar{z})$ of an operator $F$ 
is then the transform of the function
$\tilde{f}$ using the kernel function $\omega$. 

The operator $T(g)$ acts as follows:
\begin{eqnarray}
\langle z | T(g) | j,n \rangle &=& T(g) f_n(z) = (\beta z + \bar{\alpha})^{2j}
f_n \left( \frac{\alpha z - \bar{\beta}}{\beta z + \bar{\alpha}} \right)
\end{eqnarray}
We can then compute the symbol of the operator $T(g)$:
\begin{eqnarray}
\frac{ \langle \bar{z} | T(g) | \bar{z} \rangle }{   \langle \bar{z} | \bar{z} \rangle         }
&=& (1+ z \bar{z})^{-2j}
\sum_{n=0}^{2j}  \langle \bar{z} | T(g) |j,n \rangle \langle j,n | \bar{z}
\rangle
\nonumber \\
&=& (1+ z \bar{z})^{-2j} (\bar{\alpha} + \beta \bar{z} - \bar{\beta} z + \alpha z \bar{z})^{2j}
\end{eqnarray}
This indeed agrees with the kernel (46) in \cite{Grosse:1993uq}.

More crucial for comparison with previous sections is the subalgebra
of star polynomials identified in \cite{Grosse:1993uq}. We define a
set of symbols 
which are obtained by
acting with the left-invariant derivative on the symbol of $T(g)$ at
the identity:
\begin{eqnarray}
\phi_j(\bar{z} ) &=& \imath (X_j \omega (e, \bar{z}) )
\end{eqnarray}
The operator associated to this symbol is thus $(X_j T)(e)$. 

It should be noted that these operators are, from the perspective of the group
$G$,
nothing but the generators of the representation $T(g)$.
Thus, the product naturally satisfies the Lie
algebra relations (see \cite{Grosse:1993uq}).
%
%
Since these are constant $(2j+1) \times (2j+1)$
matrices, they naturally form a (small) subalgebra of the algebra of
symbols.
It is clear (via the definition in
terms of matrices) that the star product closes within this space of
symbols. It is this subalgebra that can be identified as the fuzzy
sphere \cite{Grosse:1993uq}. Thus, the fuzzy sphere coincides with the (subalgebra of the
symbols appearing in the) Berezin quantization
of the two-sphere.


\subsection*{Algebraic limits}
Now that we established that the fuzzy sphere is the Berezin quantization
of the sphere, we can use the mathematical studies of Berezin quantization
to define classical limits of the fuzzy sphere product.
The sense in which the limiting procedure from the 
Berezin quantization of the manifold to the classical
manifold is well-behaved has been the subject of many investigations.
A mathematically precise treatment is in \cite{Rieffel:2001qw} and is
based on a quantum Gromov-Hausdorff distance defined for $C^\ast$
algebras which is based on the limits defined in
\cite{Landsman:1998jw,Bordemann:1993zv}.  A good physical
understanding for these limits follows from the work of
\cite{Lieb,Simon:1980yq}, where the classical free energy (or
other observables) 
are recovered as the large
dimension limit of a quantum spin system, by proving two inequalities
for the free energy (or other observables), thus sandwiching a
classical observable in the semi-classical $h \to 0$ limit. We refer
to \cite{Rieffel:2001qw} for a recent discussion of these limits and
further developments. We conclude that the classical limit of the Berezin
sphere is well-behaved in this sense.

\subsection*{The relation to star products}
\label{star}
In this section, we want to investigate the link of our construction of
associative products to deformation quantization.
Note that in Berezin quantization it is proved that the associative product
of functions consists of a first term that is the ordinary product, a
second term that is the deformation in the direction of the Poisson
structure, and so on (see e.g. \cite{Perelomov:1986tf}).  Thus, it is natural to
think of the function algebra appearing after Berezin quantization as
a deformation of the ordinary function algebra on the two-sphere.
Here, we want to recall rigorous mathematical results that connect Berezin
quantization to geometric quantization and deformation
quantization\footnote{We are aware that many publications have analyzed these
  links (see e.g.\cite{Freidel:2001kb} for a recent discussion), and that the
  link between these domains is sometimes considered as well-known, but we
  have not been able to locate a precise formulation of the notion of
  equivalence, the limiting procedure to be employed, and the proof of
  equivalence, except in the mathematics literature that we cite below. For
  solid results in this direction, but with a slightly different limiting
  procedure, see \cite{Bordemann:1993zv}.}.

We first recall from \cite{Rawnsley:1990tj} that 
the space of Berezin symbols includes the space of
quantizable functions in geometric quantization.  For a compact
manifold, we recall that the space of symbols is a finite dimensional
subspace of the space of smooth functions on the manifold $M$. 
In fact, in the case of the sphere, it exactly
coincides with the algebra of Lie algebra generators evaluated in the
coherent state basis.  It is moreover shown in \cite{Rawnsley:1990tj}
quite generally, that the Berezin quantization of a space using a line
bundle $L$ to the power $N$ is contained in the quantization of the
space with line bundle $L^{N+1}$. For the sphere, this is simply 
 the statement that the algebra of generators for the $N$
dimensional representation is included in the algebra of generators
for the $N+1$ dimensional representation.

Crucially, the paper \cite{Rawnsley:1990tj} shows that in the large
$N$ limit, the algebra of symbols, which is the union of the algebra
of symbols at any $N$ is dense in the subalgebra of continuous
functions on $M$. In particular, the algebra separates any two points
in $M$.

Furthermore, the product of symbols of operators can be interpreted,
for any given $N$ as an associative product of functions on the
manifold $M$ -- we call this product the $N$-product.  In a follow-up
paper \cite{RCGII} it is shown that there is a formal differential
star product with deformation parameter $\nu=i/(4 \pi N)$ which
coincides with the asymptotic expansion of the $N$-product for any
compact generalized flag manifold. The differential operators arising
in the expansion depend only on the geometry of the manifold $M$, and
the remainder term goes to zero like a power of $N$ (where the power
depends on the number of terms in the expansion).  Moreover, for any
hermitian symmetric space, the formal star product is convergent.

Thus, the results of \cite{Rawnsley:1990tj,RCGII} show in a precise
way that the star $N$-product arising from Berezin quantization of the
two-sphere (amongst other manifolds) asymptotes to a formal
$*$-product on the two-sphere in the large $N$ limit.
(This completes a proof sketched in \cite{Freidel:2001kb}.)
We must note here that the star product so obtained is formal. The formal
series defining the multiplication of functions does not have a fixed radius
of convergence for a set of functions which is dense in the set of continuous
functions on the sphere. 

\section{The $_3F_2 (1)$ hypergeometric functions and 3j symbol
  transformations}
\label{whipple}
In this appendix we will recall 
relations between $_3F_2 (1)$
hypergeometric functions \cite{Whipple26,Whipple26b,Whipple27,Slater}
 and use these relations to show how one can
determine the symmetry properties of $SU(2)$ or $SU(1,1)$ 3j symbols.

\subsection*{Whipple functions}
We first introduce Whipple's notation, which is a very compact way to
write down the various transformations relating different $_3F_2 (1)$
functions\footnote{By an $_3F_2 (1)$ function we mean a
  hypergeometric function of five complex arguments of the kind $_3F_2
  (a_1,a_2,a_3;b_1,b_2;1)$.}. In this notation, one has six complex
parameters $r_i$, $i=0,1,...,5$, which obey the following condition:
\begin{eqnarray}
\label{sumr}
r_0+r_1+r_2+r_3+r_4+r_5 = 0
\end{eqnarray}
therefore leaving us with five degrees of freedom. One
then defines the following variables:
\begin{eqnarray}
\alpha_{lmn} &=& \frac{1}{2}+r_l+r_m+r_n \nonumber \\
\beta_{lm} &=& 1+r_l-r_m
\end{eqnarray}
Some simple remarks are in order: $\alpha_{lmn}$ is completely
symmetric, while $\beta_{lm} = 2-\beta_{ml}$. Also, note that due to
equation $(\ref{sumr})$, we have $\alpha_{lmn} = 1-\alpha_{ijk}$ where the
indices $i,j,k$ are
all different from $l,m,n$.

Using this notation, the (Thomae-)Whipple functions are:
\begin{eqnarray}
F_p (l;m,n) &=& \frac{1}{\Gamma(\alpha_{ghj}, \beta_{ml}, \beta_{nl})}
\ {_3}F_2 \left( \begin{array}{c}
\alpha_{gmn}, \ \alpha_{hmn}, \ \alpha_{jmn} \\
\beta_{ml} \ , \ \ \beta_{nl}
\end{array} ;1 \right) \nonumber \\
F_n (l;m,n) &=& \frac{1}{\Gamma(\alpha_{lmn}, \beta_{lm}, \beta_{ln})}
\ {_3}F_2 \left( \begin{array}{c}
\alpha_{lgh}, \  \alpha_{lgj}, \ \alpha_{lhj} \\
\beta_{lm} \ , \ \  \beta_{ln}
\end{array} ;1 \right)
\end{eqnarray}
where $g$, $h$ and $j$ are indices all different from $l$, $m$ and
$n$. Note that any $F_n (l;m,n)$ function is obtained from the
corresponding $F_p (l;m,n)$ function by changing the signs of all the
parameters $r_i$. The function $F_p (l;m,n)$ is well-defined if $\Re
\alpha_{ghj} >0$ and $F_n (l;m,n)$ is well-defined if $\Re
\alpha_{lmn} >0$. This is related to the fact that the defining series
for $_3F_2 (a_1,a_2,a_3;b_1,b_2;1)$ is well-defined iff $s =
b_1+b_2-a_1-a_2-a_3$ is such that $\Re s >0$, assuming that no
argument is a negative integer. Things are a little different if at
least one $a_i$ is a negative integer, since in this case $_3F_2
(a_1,a_2,a_3;b_1,b_2;1)$ can be expressed as a finite sum and there is
no more requirement on s.  There is however a complication if some
$a_i$'s and some $b_j$'s are negative integers at the same time, for
in this case there is no clear limiting procedure that would allow us
to define $_3F_2 (1)$.  Assuming that $a_{i_0} = -N$ is the largest of
all negative integers $a_i$ and that any $b_j$ that is a negative
integer is smaller than $-N$, a reasonable definition is:
\begin{eqnarray}
_3F_2 \left( \begin{array}{ccc}
a_1, & a_2, & a_3 \\
b_1, & b_2 &
\end{array} \ ; 1 \right)
= \sum_{n=0}^N \ \frac{(a_1)_n (a_2)_n (a_3)_n}{(b_1)_n (b_2)_n} \
\frac{1}{n!}
\end{eqnarray}
i.e we have truncated the infinite sum. We have used the Pochammer symbol
$(a)_n = \frac{\Gamma(a+n)}{\Gamma(a)}$.

\subsection*{Two-term and three-term relations}
Relations between $_3F_2 (1)$ functions are standard and can be
elegantly written in Whipple's notation. There are two kinds of such
relations. The first one consists of two-term relations:
\begin{eqnarray}
F_p (l;m,n) &=& F_p (l;m^{\prime},n^{\prime}) \nonumber \\
F_n (l;m,n) &=& F_n (l;m^{\prime},n^{\prime})
\end{eqnarray}
i.e they are simply equivalent to the statement that both $F_p
(l;m,n)$ and $F_n (l;m,n)$ actually do not depend on $m$ and $n$
(therefore we will denote them by $F_p (l)$ and $F_n (l)$). The
second kind of relations are three-term relations:
\begin{eqnarray}
\label{3Fpn}
\frac{\sin{\pi \beta_{23}}}{\pi \Gamma(\alpha_{023})} F_p (0) &=& \frac{F_n
  (2)}{\Gamma(\alpha_{134}, \alpha_{135}, \alpha_{345})} - \frac{F_n
  (3)}{\Gamma(\alpha_{124}, \alpha_{125}, \alpha_{245})}
\\
\frac{\sin{\pi \beta_{32}}}{\pi \Gamma(\alpha_{145})}  F_n (0) &=& \frac{F_p
  (2)}{\Gamma(\alpha_{012}, \alpha_{024}, \alpha_{025})} - \frac{F_p
  (3)}{\Gamma(\alpha_{013}, \alpha_{034}, \alpha_{035})}
\nonumber \\
\frac{\sin{\pi \beta_{45}} \ F_p (0)}{\Gamma(\alpha_{012}, \alpha_{013},
  \alpha_{023})} &=& - \frac{\sin{\pi
    \beta_{50}} \ F_p (4)}{\Gamma(\alpha_{124}, \alpha_{134}, \alpha_{234})}
- \frac{\sin{\pi \beta_{04}} \ F_p (5)}{\Gamma(\alpha_{125},
  \alpha_{135}, \alpha_{235})}
\nonumber \\
\frac{\sin{\pi \beta_{54}} \ F_n (0)}{\Gamma(\alpha_{145}, \alpha_{245},
  \alpha_{345})} &=& - \frac{\sin{\pi
    \beta_{05}} \ F_n (4)}{\Gamma(\alpha_{015}, \alpha_{025}, \alpha_{035})}
- \frac{\sin{\pi \beta_{40}} \ F_n (5)}{\Gamma(\alpha_{014},
  \alpha_{024}, \alpha_{034})} \nonumber
\end{eqnarray}
\begin{eqnarray}
\frac{F_p (0)}{\Gamma(\alpha_{012}, \alpha_{013}, \alpha_{023},
  \alpha_{014}, \alpha_{024}, \alpha_{034})} &=& K_0 F_p
  (5)- \frac{\sin{\beta_{05}} \ F_n (0)}{\Gamma(\alpha_{123},
    \alpha_{124}, \alpha_{134}, \alpha_{234})}
\nonumber \\
\frac{F_n (0)}{\Gamma(\alpha_{125}, \alpha_{135}, \alpha_{235},
  \alpha_{145}, \alpha_{245}, \alpha_{345})} &=& K_0 F_n
  (5)- \frac{\sin{\beta_{50}} \ F_p (0)}{\Gamma(\alpha_{015},
    \alpha_{025}, \alpha_{035}, \alpha_{045})}
\nonumber
\end{eqnarray}
where $K_0 = \frac{1}{\pi^3} \left( \sin{\pi \alpha_{145}} \sin{\pi
    \alpha_{245}} \sin{\pi \alpha_{345}} + \sin{\pi \alpha_{123}}
  \sin{\pi \beta_{40}} \sin{\pi \beta_{50}} \right)$.  These six
identities, up to permutation of indices, give us 120 independent
relations between $_3F_2 (1)$ functions.
These three-term relations may reduce to two-term relations when one
or more $\alpha_{lmn}$ is a negative integer, as we will see below.

\subsection*{An integer limit of Whipple's relations}
\label{integer}
Let us proceed to find Whipple's relations in the specific case that is
needed in order to understand transformations of the 3j symbols,
whether we consider $SU(2)$ or $SU(1,1)$. In these cases the arguments
of the hypergeometric functions used in the expressions of the 3j
symbols are all integer. Since Whipple's relations are valid in the
generic case of complex parameters, relations in the case of integer
parameters should be obtained from a limiting procedure. With no loss
of generality, we are free to choose:
\begin{eqnarray}
\alpha_{145} &=& -n + \epsilon \ , \hspace{1.2cm}  \alpha_{014} = -n_2 +
\epsilon_2 \nonumber \\
\alpha_{245} &=& -n'_1 + \epsilon'_1 \ , \hspace{1cm} \alpha_{015} =
-n'_2 + \epsilon'_2 \\
\alpha_{345} &=& -n_1 + \epsilon_1 \nonumber
\end{eqnarray}
where $\epsilon, \epsilon_1, \epsilon'_1, \epsilon_2, \epsilon'_2$ are real
and tend to zero.  The behavior of all other parameters $\alpha_{lmn}$,
$\beta_{lm}$ is then fixed.

One important remark is that, though the $\epsilon$'s may be anything a
priori, the limiting procedure must 
satisfy some consistency conditions.
To be more precise, the fact that all 3j symbols are real and that all
ratios of 3j symbols, for $SU(2)$ on one hand, and for $SU(1,1)$ on the other
hand, are of modulus one implies that:
\begin{eqnarray}
|\epsilon| = |\epsilon_1| = |\epsilon'_1| = |\epsilon_2| =
|\epsilon'_2|
\end{eqnarray}
and that the relative signs between $\epsilon$'s must also be of the
following kind only (up to a global sign that we use to fix $\epsilon >0$):
\begin{eqnarray}
\begin{array}{ccccc}
\epsilon & \epsilon_1 & \epsilon'_1 & \epsilon_2 & \epsilon'_2 \\
+ & + & + & + & - \\
+ & + & + & - & + \\
+ & + & - & + & + \\
+ & - & + & + & + \\
+ & + & + & - & - \\
+ & - & - & + & +
\end{array}
\end{eqnarray}
Now that we know what kind of limit we can consider, and that results
should not depend on which of the six possible limits we take, it is
possible to find the following relations resulting from equations
$(\ref{3Fpn})$ in this limit:
\begin{eqnarray}
\label{integerlimitw}
F_p (0)&=& (-1)^{\beta_{04}-1} \Gamma \left( \begin{array}{ccc}
\alpha_{012} & \alpha_{013} & \alpha_{023} \\
\alpha_{124} & \alpha_{134} & \alpha_{234}
\end{array} \right) F_p (4) \\
&=& (-1)^{\beta_{05}-1} \Gamma \left( \begin{array}{ccc}
\alpha_{012} & \alpha_{013} & \alpha_{023} \\
\alpha_{125} & \alpha_{135} & \alpha_{235}
\end{array} \right) F_p (5) \nonumber \\
F_n (1) &=& (-1)^{\beta_{12}-1} \Gamma \left( \begin{array}{ccc}
\alpha_{023} & \alpha_{234} & \alpha_{235} \\
\alpha_{013} & \alpha_{134} & \alpha_{135}
\end{array} \right) F_n (2) \nonumber \\
&=& (-1)^{\beta_{13}-1} \Gamma \left( \begin{array}{ccc}
\alpha_{023} & \alpha_{234} & \alpha_{235} \\
\alpha_{012} & \alpha_{124} & \alpha_{125}
\end{array} \right) F_n (3) \nonumber \\
F_p (2) &=& - \Gamma \left( \begin{array}{ccc}
\alpha_{023} & \alpha_{234} & \alpha_{235} \\
\alpha_{013} & \alpha_{125} & \alpha_{135}
\end{array} \right) F_p (1) \nonumber \\
&=& - \Gamma \left( \begin{array}{ccc}
\alpha_{012} & \alpha_{124} & \alpha_{125} \\
\alpha_{013} & \alpha_{134} & \alpha_{135}
\end{array} \right) F_p (3) \nonumber \\
F_n (4) &=& \Gamma \left( \begin{array}{ccc}
\alpha_{012} & \alpha_{013} & \alpha_{023} \\
\alpha_{124} & \alpha_{134} & \alpha_{234}
\end{array} \right) F_n (0) \nonumber \\
&=& \Gamma \left( \begin{array}{ccc}
\alpha_{125} & \alpha_{135} & \alpha_{235} \\
\alpha_{124} & \alpha_{134} & \alpha_{234}
\end{array} \right) F_n (5) \nonumber \\
F_p (0) &=& (-1)^{\alpha_{145}} \Gamma \left( \begin{array}{cc}
\alpha_{012} & \alpha_{013} \\
\alpha_{234} & \alpha_{235}
\end{array} \right) F_n (1) \nonumber \\
\epsilon^2 F_p (0) &\sim& (-1)^{\beta_{45}-1} \frac{\Gamma
  (\alpha_{013})}{\Gamma (\alpha_{123}, \alpha_{124}, \alpha_{125},
  \alpha_{234}, \alpha_{235})} \ F_p (2) \nonumber \\
\epsilon^2 F_n (1) &\sim& (-1)^{\beta_{23}} \frac{\Gamma
  (\alpha_{234})}{\Gamma (\alpha_{012}, \alpha_{013}, \alpha_{123},
  \alpha_{125}, \alpha_{135})} \ F_n (4) \nonumber 
\end{eqnarray}
Note that, while some Whipple functions are finite, like $F_p (0)$,
others tend to zero in the integer limit.

From the above relations it is possible to find the
symmetry relations of 3j symbols, by choosing appropriately the
parameters $\alpha_{lmn}$, $\beta_{lm}$. There is however one subtlety
that arises in some cases: $\Gamma$ function prefactors may contribute
to the overall sign that relate any 3j symbol to any other 3j symbol.
Recall for instance the expression of the $SU(1,1)$ 3j symbol that was
given in equation $(\ref{3jsu11})$. It is not clear whether one
should put a given $\Gamma$ function (say $\Gamma (m_1+j_1)$ for
instance) inside or outside the square root, since this does not
matter for the $+++$ 3j symbol. It would however matter if one would
try to compute some other 3j symbol, like $---$.

This last subtlety can be solved by computing explicitly a 3j symbol for
$SU(1,1)$ involving at least one $\mathcal{D}^-$ representation.
This can be done using a result from
\cite{Davids:2000kz}. The 3j symbol given in \cite{Davids:2000kz} is
related to our formula $(\ref{3jsu11})$ by:
\begin{align}
& \left[ \begin{array}{ccc}
j_1 & j_2 & j \\
m_1 & m_2 & -m
\end{array} \right]^{+++}_{D.} = (-1)^{j+j_1-m_2+1}
\left[ \begin{array}{ccc}
j_2 & j_1 & j \\
m_2 & m_1 & -m
\end{array} \right]^{+++}_{us} \\
& = (-1)^{j_1-m_1}  \sqrt{\frac{(m_1+j_1-1)!
    (m_1-j_1)! (m_2+j_2-1)! (m_2-j_2)!}{(m-j)! (m+j-1)!}}
\nonumber \\
& \times \sqrt{(j+j_1+j_2-2)! (j+j_1-j_2-1)! (j-j_1+j_2-1)!
  (j-j_1-j_2)!}
\nonumber \\
& \times \frac{1}{(m_1+j_1-1)! (m_1-j_1)!} \cdot
    \frac{1}{(j+j_2-m_1-1)! (j-j_2-m_1)!} \nonumber \\
& \times \ _3F_2 \left(
\begin{array}{c}
j-m, \ \ 1-m_1-j_1, \ \ j_1-m_1 \\
j+j_2-m_1, \ \ j-j_2-m_1+1
\end{array} ; 1 \right) \nonumber
\end{align}
It satisfies\footnote{There should actually be a factor $\text{sign}
  (\epsilon \epsilon_1 \epsilon_1^{\prime})$ in the second equality.
  We take this sign to be $+1$.}:
\begin{eqnarray}
\left[ \begin{array}{ccc}
j_1 & j_2 & j \\
m_1 & m_2 & -m
\end{array} \right]^{+++}_{D.}
&=& (-1)^{j-j_1-j_2}
\left[ \begin{array}{ccc}
j_2 & j_1 & j \\
m_2 & m_1 & -m
\end{array} \right]^{+++}_{D.} \nonumber \\
&=&
\left[ \begin{array}{ccc}
j_1 & j_2 & j \\
-m_1 & -m_2 & m
\end{array} \right]^{+++}_{D.}
\end{eqnarray}
This calculation forces us to place particular $\Gamma$ functions
(factorials) outside of the square root according to the above formula
for the 3j symbols. This resolves the ambiguity in the relative sign
of the 3j symbols as we analytically continue. Other symmetry
relations of the $SU(1,1)$ 3j symbols follow directly from equations
$(\ref{integerlimitw})$.

\end{document}